\documentclass[useAMS,usenatbib]{mn2e}
\usepackage{graphicx}
\usepackage{graphics}
\usepackage{amsmath}
\usepackage{txfonts}
\usepackage{rotating}
\usepackage{lscape}
\usepackage{textcomp}
\usepackage{amssymb}

\title[Variability selected AGN in GOODS-S]{The SEDs, Host Galaxies and Environments of Variability selected AGN in GOODS-S}
\author[Villforth et
al.]{ Carolin Villforth$^{1,2}$\thanks{E-mail:villforth@astro.ufl.edu},
Vicki Sarajedini$^{1}$ and
Anton Koekemoer$^{2}$
\\
$^{1}$Department of Astronomy, University of Florida, 32611 Gainesville, FL,
USA\\
$^{2}$Space Telescope Science Institute, 21218 Baltimore, MD, USA
\\ }
\begin{document}

\date{Accepted July 17th 2012}

\pagerange{\pageref{firstpage}--\pageref{lastpage}} \pubyear{2012}

\maketitle

\label{firstpage}

\begin{abstract}
Variability selection has been proposed as a powerful tool for identifying both low-luminosity AGN and those with unusual SEDs. However, a systematic study of sources selected in such a way has been lacking. In this paper, we present the multi-wavelength properties of the variability selected AGN in GOODS South. We demonstrate that variability selection indeed reliably identifies AGN, predominantly of low luminosity. We find contamination from stars as well as a very small sample of sources that show no sign of AGN activity, their number is consistent with the expected false positive rate. We also study the host galaxies and environments of the AGN in the sample.  Disturbed host morphologies are relatively common. The host galaxies span a wide range in the level of ongoing star-formation. However, massive star-bursts are only present in the hosts of the most luminous AGN in the sample. There is no clear environmental preference for the AGN sample in general but we find that the most luminous AGN on average avoid dense regions while some low-luminosity AGN hosted by late-type galaxies are found near the centres of groups. AGN in our sample have closer nearest neighbours than the general galaxy population. We find no indications that major mergers are a dominant triggering process for the moderate to low luminosity AGN in this sample. The environments and host galaxy properties instead suggest secular processes, in particular tidal processes at first passage and minor mergers, as likely triggers for the objects studied. This study demonstrates the strength of variability selection for AGN and gives first hints at possibly triggering mechanisms for high-redshift low luminosity AGN. 
\end{abstract}


\section{Introduction}
\label{S:intro}

While our understanding of the physics of AGN has grown considerably over the last decades, an important questions about AGN remains unanswered until today: How are AGN triggered? This question has become of interest also for a wider field of astronomy research when it became clear that AGN likely play an important role in the evolution of galaxies. The masses of the central black holes in nearby galaxies correlate surprisingly well with the properties of their hosts \citep{gebhardt_relationship_2000,graham_correlation_2001,ferrarese_fundamental_2000,novak_correlations_2006,graham_expanded_2010} It has been argued that these correlations imply a causal connection between AGN and their hosts, or at least the galactic bulges of their hosts \citep[e.g.][]{silk_quasars_1998}. However, other authors have argued that the data does not necessarily imply a casual connection \citep{peng_how_2007,jahnke_non-causal_2010}, but can be explained by progression to the mean during mergers throughout cosmic time. Also, the fact that both star formation and AGN activity peak around a redshift of two has been used as an argument for a causal connection between AGN and galaxies \citep[e.g.][]{boyle_cosmological_1998,merloni_tracing_2004}.

Theoreticians have suggested several mechanisms in which AGN activity and galaxy evolution could be connected \citep[e.g.][]{silk_quasars_1998,sanders_luminous_1996,hopkins_cosmological_2008}. Major mergers of galaxies have been the main mechanisms discussed \citep{sanders_luminous_1996,silk_quasars_1998,hopkins_characteristic_2009}, in particular for the most luminous quasars. The evidence for major mergers as triggers for AGN activity has been mixed \citep[e.g.][and references therein]{canalizo_quasi-stellar_2001,hopkins_cosmological_2008,kocevski_candels:_2011,cisternas_bulk_2011,schawinski_heavily_2012} and latest studies seem to imply that earlier findings are mostly due to selection effects \citep{kocevski_candels:_2011,cisternas_bulk_2011}. Recent studies have also emphasized that other mechanisms might play a role in triggering AGN activity in particular for low luminosity AGN and heavily obscured systems \citep{hopkins_characteristic_2009,bournaud_black_2011}.  These studies have emphasized that to properly understand AGN activity, one needs to study a wide range of AGN types, both in luminosity and obscuration. 

Host galaxy studies commonly focus either on low redshift systems \citep[e.g.][]{canalizo_quasi-stellar_2001,veilleux_deep_2009} or high luminosity quasars at high redshift \citep[e.g.][]{villforth_quasar_2008,schramm_host_2008,kocevski_candels:_2011}. A part of the parameter space that has been poorly explored are low luminosity AGN at high redshift, this is in particular due to the difficulty in identifying high-redshift low luminosity AGN. 

Variability selection has been used as a powerful tool for selecting AGN, in particular due to the emergence of large surveys in the last decades \citep{schmidt_selecting_2010,macleod_quasar_2011}. In particular, applying it to high-resolution HST data has been recognized as a valuable tool for selecting AGN at low luminosities up to high redshift \citep[e.g.][]{sarajedini_v-band_2003,villforth_new_2010,sarajedini_variability_2011}. In a recent paper, \citet{villforth_new_2010} published a catalogue of variability selected AGN in the GOODS Fields, containing a total of 155 variable sources, 88 of those located in GOODS-South. The objects found in this study are predominantly of Seyfert luminosities and have dominant host galaxies. In this paper, we aim to analyse  the multi-wavelength properties of the 88 variability selected AGN candidates. We will summarize the properties of variability selected AGN and re-evaluate the issue of contamination from stars and false positives. We will also analyse this particular sample to asses questions concerning the likely triggering mechanisms and environments of lower luminosity AGN at higher redshift. This study will therefore explore a part of AGN parameter space that has been neglected so far and analyze the strength and weaknesses of variability selection.

The paper is structured as follows: Section 2 recaps the sample selection from \citet{villforth_new_2010} and presents a short summary of data used in this study. Section 3 analyses the SEDs and general properties of the variability selected sources and discusses contamination by both stars and false positives. Host galaxies are analysed in Section 4. Section 5 discusses environments of variability selected AGN, followed by a discussion of the results in Section 6 and summary and conclusions in Section 7. The cosmology used is $H_{0}=70\textrm{km s}^{-1}\textrm{Mpc}^{-1},\Omega_{\Lambda}=0.7, \Omega_{m}=0.3$. Throughout the paper, we use AB magnitudes.

\section{Data}
\label{S:data_main}

\subsection{Sample}

In this study, we will use the variability selected sample from a previous study by the authors, details of the method used can be found in \citet{villforth_new_2010}, but the basic methods will be discussed here shortly.

Variability selection is performed on five epochs of GOODS data taken with the Advanced Camera for Surveys (ACS) Wide Field Channel (WFC) aboard HST in the F850LP (z) band. Data reduction is performed using MultiDrizzle \citep{koekemoer_multidrizzle:_2002,koekemoer_candels:_2011}. The five epochs were separated by about 6 weeks in time, overall spanning about 7 months. An initial selection was performed to reject objects with SN $<$ 20 in the combined five epoch data. Additionally, only objects with good data in all five epochs were included.

The flux measurements were done by simple aperture photometry, tests were done with a range of aperture sizes, but finally an aperture radius of 0.36 " was chosen. Errors were determined using weight maps and corrected using the full sample. Visual inspection was performed to check for diffraction spikes and affected objects were discarded. The final sample was then selected using a $C$ statistic. A significance level of 99.99\% was chosen to limit contamination by false positives, additionally a more rigorous 'clean' catalogue (significance level 99.99\%) which is expected to contain only one false positive is provided.

All sources where then cross-checked with a number of catalogues and databases to reject stars. After this, in the GOODS-S field 88 objects were variability selected, of which 14 where identified as stars, leaving 74 AGN candidates. We expect six sources to be false positives.

\subsection{Short Description of Datasets Used}
\label{S:data}

In this study, we combine three different data sets covering the GOODS South field: the photometric redshift and multi-waveband optical-IR catalogue used in \citet{dahlen_detailed_2010}, the CDFS 4 Ms X-ray catalogue from \citet{xue_chandra_2011} as well as the CDFS VLA 20cm and 6cm from \citet{kellermann_vla_2008}.

We use photometric redshifts and multi-wavelength data from \citet{dahlen_detailed_2010}. This dataset includes photometric data for the following bands: VLT/VIMOS (U-band), HST/ACS (F435W, F606W, F775W, and F850LP bands), VLT/ISAAC (J-, H-, and Ks-bands) as well as four Spitzer/IRAC channels (3.6, 4.5, 5.8, and 8.0$\mu$m).

The spatial resolution for the data at different wavelength are as follows: 0.11" for the HST data, 0.35-0.65" for the ground-based IR data, 0.8" for the U band data and 1.7-1.9" for the IRAC data. The resolution for the Chandra data is about 0.5" with a typical spatial uncertainty of 0.4". Finally, the VLA data have a typical resolution of about 3.5".

We also use the CDFS 4Ms catalogue from \citet{xue_chandra_2011}, this catalogue considerably improved the sensitivity from existing X-ray catalogues \citep{luo_chandra_2008} to $3.2 \times 10^{-17} \textrm{erg} \textrm{cm}^{-2} \textrm{s}^{-1} , 9.1 \times 10^{-18} \textrm{erg} \textrm{cm}^{-2} \textrm{s}^{-1} and 5.5 \times 10^{-17} \textrm{erg} \textrm{cm}^{-2} \textrm{s}^{-1}$ for the full, soft, and hard bands, respectively. Radio data at 1.4GhZ (20cm) and 4.8GhZ (6cm) is taken from \citet{kellermann_vla_2008}, with a rms per beam between 9-43 $\mu Jy$ for the 20cm band (lowest rms/beam is reached in the center of the field).

Both X-ray and radio catalogues are matched to the $z$-band variability selected catalogue in a simple radius matching technique. Matching radii correspond to the typical resolution of the data. The matching procedure is analysed to verify that there are no offsets between the coordinate systems. For this purpose, we analyse the distributions of distances in both right ascension and declination and check for offsets from zero. These offsets are then corrected for and the matching process is iterated. We find a small offset of about 0.1" in both right ascension and declination for the 4Ms data only. Correcting for it however does not significantly change the number of matches. No offset is found in the radio data matching.

\section{Spectral Energy Distributions and General Properties of Variability Selected AGN}
\label{S:sample}

With a 99.99\% significance level, 88 variable objects were found in GOODS South, after rejecting stars, 74 variability selected AGN remain, of those sources, 21 had reported spectroscopic redshifts \citet{villforth_new_2010}. 21 additional sources have spectroscopic redshifts reported in the current paper, yielding a total of 42 sources
with spectroscopic data. Three of the 88 initial sources do not show up in the multiwavelength catalogue and are thus discarded, leaving 71 objects studied here. All references for the spectroscopic redshifts can be found in Table \ref{T:AGN} in the Appendix. Spectroscopic redshifts range from 0.0459 to as high as 3.7072. Adding the photometric redshift data from \citet{dahlen_detailed_2010} gives us redshift estimates for the rest of the sources. The photometric redshifts range from 0.04 to 4.34. The mean and median of the redshift distribution are 0.94 and 0.6.

\subsection{SED Fitting Procedure}
\label{S:sed}

We make use of the wide wavelength coverage in the optical-NIR available for GOODS South. As a comparison, we use the SWIRE Template Library \footnote{$http://www.iasf-milano.inaf.it/~polletta/templates/swire\_templates.html$}, which provides templates over a wide wavelength range for normal quiescent elliptical and spiral galaxies, starburst galaxies, quasars as well as mixed
starburst and AGN objects. The templates are based on data from \cite{silva_modeling_1998,gregg_reddest_2002,berta_spatially-resolved_2003,hatziminaoglou_sloan_2005,polletta_chandra_2006}. Additionally, we use the Kurucz 1993 stellar models \citep{kurucz_model_1979} \footnote{Downloaded from http://www.stsci.edu/hst/observatory/cdbs/k93models.html}. SED Fitting is performed using a simply least square minimization technique. In a first step, stellar SEDs at a redshift of zero are fit to identify stellar contaminants, this is discussed in Section \ref{S:stars}. AGN/Galaxy combination fits as well as typical errors, caveats, possible biases and sources of systematic errors are then discussed in Section \ref{S:sed_agn}.

\subsubsection{Rejection of Stars}
\label{S:stars}

Variable stars are a possible contaminant to the variability selected AGN sample. While the GOODS-South field is located far from the galactic plane, halo stars could still be present. It is also known that many stars show variability on time scales larger than a few hours, with a majority showing variability on the scale of a few percent on week and month time scales \citep[e.g.][]{hartman_photometric_2011,bryden_kepler_2011}. While this is generally shorter than typical variability timescales of AGN, which tend to be on the order of weeks instead of hours \citep[e.g.][]{kelly_are_2009}, stars will still show detectable variability on the month time-scales sampled by the data. Therefore, initially selecting against point sources is not useful since quasars are point sources.

In \citet{villforth_new_2010}, we identified 14 objects as stars through archive searches. Using SED Fitting, we additionally identify 10 objects as stars, usually of G-M type stars. Note that a majority of these stars were identified as low redshift early type galaxies in the photometric redshift catalogues due to their SED shape. Only the mid-IR data allows to reliably distinguish those objects from faint red galaxies.

In the following analysis, we will no longer include objects identified as stars through either SED fitting or through literature search. They are still listed in Table \ref{T:Stars}. After flagging all stellar objects, a total of 61 non-stellar variability selected sources remain. We will not further consider objects identified as stars and continue with those 61 objects, which we will label AGN candidates.

\subsubsection{Combination Fits of Remaining AGN Candidates}
\label{S:sed_agn}

For the sample of 61 AGN candidates, we perform mixed SED fits. This means that we fit a mixture of AGN and galaxy SEDs, we perform these fits for a range of different input SEDs for both the AGN and galaxy components.In total 6 different AGN SEDs and 11 different galaxy SEDs where used. For each of those combinations, we vary the fractional distribution of the galaxy and AGN in the $z$ band and fit the combined SED. The best fits for each combination are then compared and the overall best fit is selected. All results of the SED Fits are provided in Table \ref{T:AGN} in the Appendix.

Out of the 61 objects, a clear best fit was found for 47 objects. For the other 14 objects, the fit was ambiguous. We could not determine the properties of those sources reliably. The sources with failed fits are predominantly faint, extended and show clear disturbed morphology. The failed fits might be due to several reasons. Due to the long time-span over which the data were taken, the AGN variability can cause excess scatter. Strong line emission could cause additional bumps in the SED and multi-wavelength photometry might be less reliable for sources with complex morphology due to the differences in resolution. Additionally, failed fits generally show erratic SEDs that might be indicative of an AGN dominated SED with strong variability. The failed fits will be discussed in more detail in Section \ref{S:failed_fits}.

The SED fits contain several sources of errors. We will first qualitatively discuss different influences and then quantify these effects to derive expected errors on determined magnitudes:

\begin{itemize}
\item \textbf{Variability:} The AGN in this sample were selected through their variability. Due to the fact that the multi-wavelength data were taken over a timespan of several years \citep{dahlen_detailed_2010}, the AGN variability causes additional scatter in the SED that are not accounted for in the photometric errors. This causes poorer fits than expected. Therefore SED fitting is especially difficult for sources that are AGN dominated, as well as for those with already large photometric errors (see also Section \ref{S:failed_fits}). A proper treatment of this effect is complicated since variability strength varies across the SED and photometric data is often taken over a wide time-span \citep{dahlen_detailed_2010}. This is beyond the scope of the current project. However, we do not expect this effect to cause systematic errors in our fits.
\item \textbf{Size of template library:} For template fitting, there is a trade-off between using a large sets of templates and therefore being able to account for subtle differences in the SEDs against using a smaller subset, resulting in less accurate fits but more robust results. Due to the complexity of the SEDs, we decide for a limited template library. A larger number of templates will cause difficulties in finding global minima in the fit.  
\item \textbf{Unclear Redshift:} Two thirds of the AGN (42/61) in this sample have spectroscopic redshifts, leaving 19 objects for which we rely on photometric data. Some of the spectroscopic redshifts are of lower quality (i.e. rely on noisy spectra or single line detections). While the photometric redshifts of sources with significant galaxy contribution are relatively reliable (see Section \ref{S:redshift}), some of the sources have very uncommon SEDs and we cannot find reliable redshift fits. While the majority of the sample has accurate enough redshifts to have little influence on the resulting fits, a few sources have SEDs for which we can not determine the proper redshift (see Section \ref{S:failed_fits} as well as Fig. \ref{F:Ind_bumpy1}).
\end{itemize}

Taking into account the factors mentioned above as well as typical uncertainties in the fit, the resulting errors in the AGN and galaxy magnitudes are typically 0.5-0.7 magnitudes for the non-dominant component and around 0.1 magnitudes for the dominant component. Errors in the percentage contribution are typically 1-5 per cent points.

\begin{figure}
\includegraphics[width=8cm]{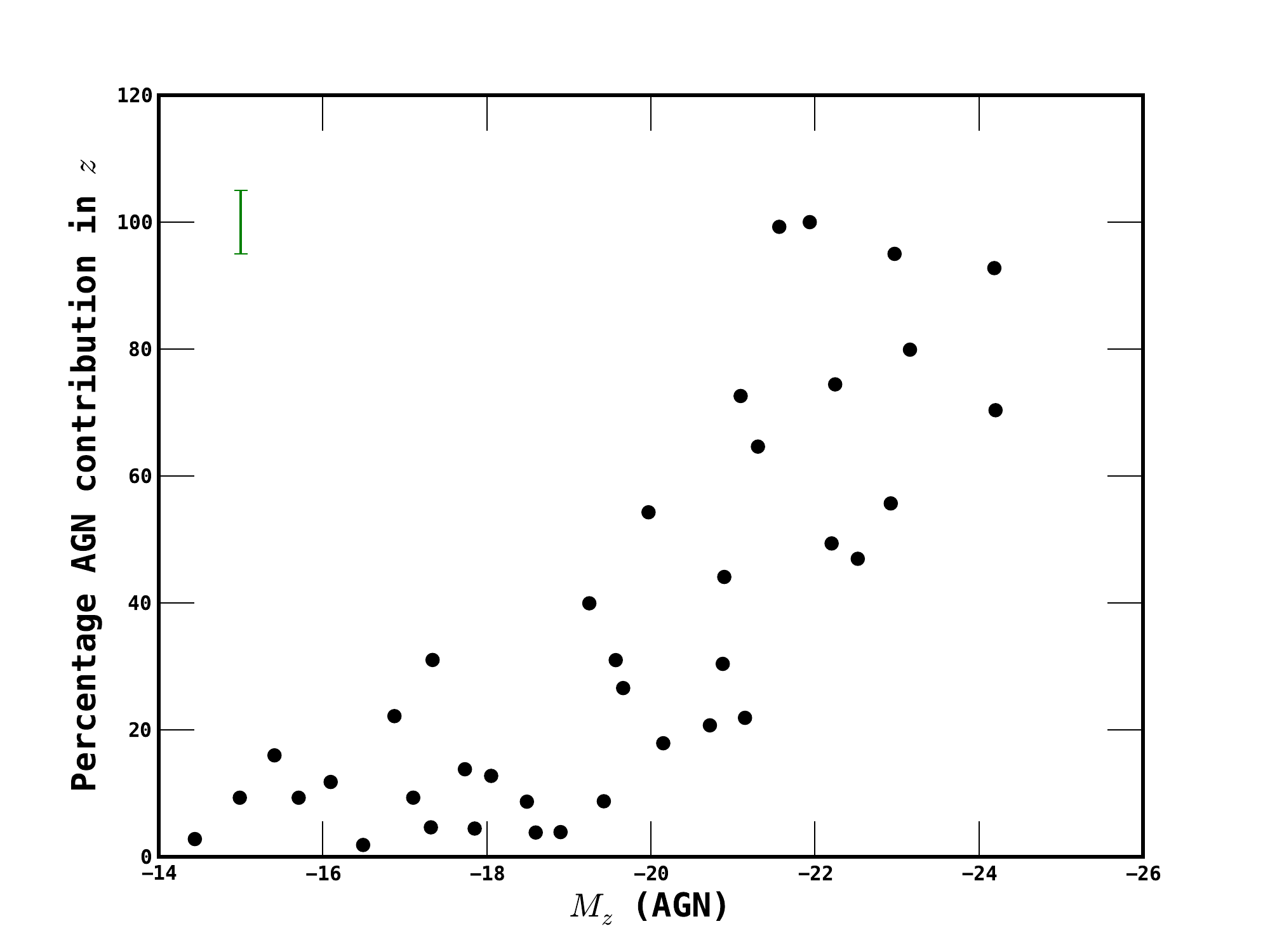}
\caption{Percentage contribution of AGN in $z$ plotted against absolute $z$-band AGN magnitude. Only sources with successful fits that have not been identified as false positives are shown. Errors in the AGN magnitudes are between 0.1 and 0.5 mags and errors in the AGN contribution are $\leq$ 5 percent points.}
\label{F:LP_AGNMags}
\end{figure}

\begin{figure}
\includegraphics[width=8cm]{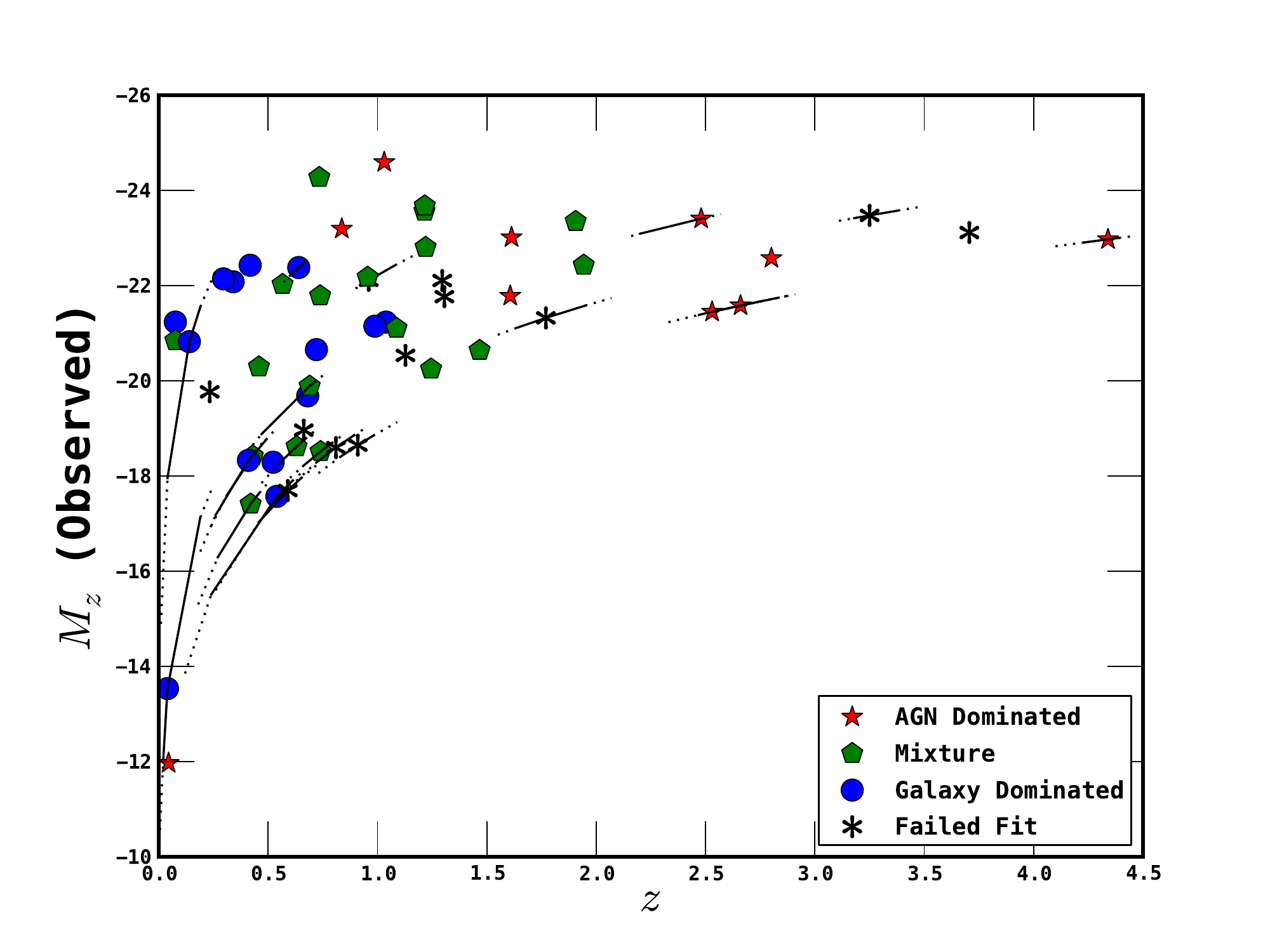}
\caption{ Observed $z$ band magnitudes versus redshift for all variability selected AGN by AGN contribution: black x: failed fit; red star: AGN dominated, green pentagons: mixture, blue circles: galaxy dominated. The magnitudes are shown for the entire object, i.e. contribution from the host galaxy is not subtracted off.}
\label{F:SED_General}
\end{figure}

To determine the actual luminosity of the AGN component, it is important to disentangle the contribution of galaxy and AGN. Fig. \ref{F:LP_AGNMags} shows the contribution of the AGN to the overall magnitude for all variability selected sources for which the fit succeeded. The AGN contribution can be as low as $\sim$5\%. In the following, we will call sources with $>$90\% AGN contribution in $z$ AGN dominated, those with $<$10\% AGN contribution galaxy dominated and all others mixtures. Fig. \ref{F:SED_General} shows the magnitudes and redshifts of AGN dominated, mixture and galaxy dominated sources. Galaxy dominated sources are only present at very low redshifts where the small aperture used in \citet{villforth_new_2010} only includes a small proportion of galaxy light and therefore makes the detection of variability of a faint source against a background galaxy easier. A majority of sources beyond redshift 0.5 are mixture objects.

\subsection{Redshift distribution of AGN candidates and accuracy of
photometric redshift}
\label{S:redshift}

\begin{figure}
\includegraphics[width=8cm]{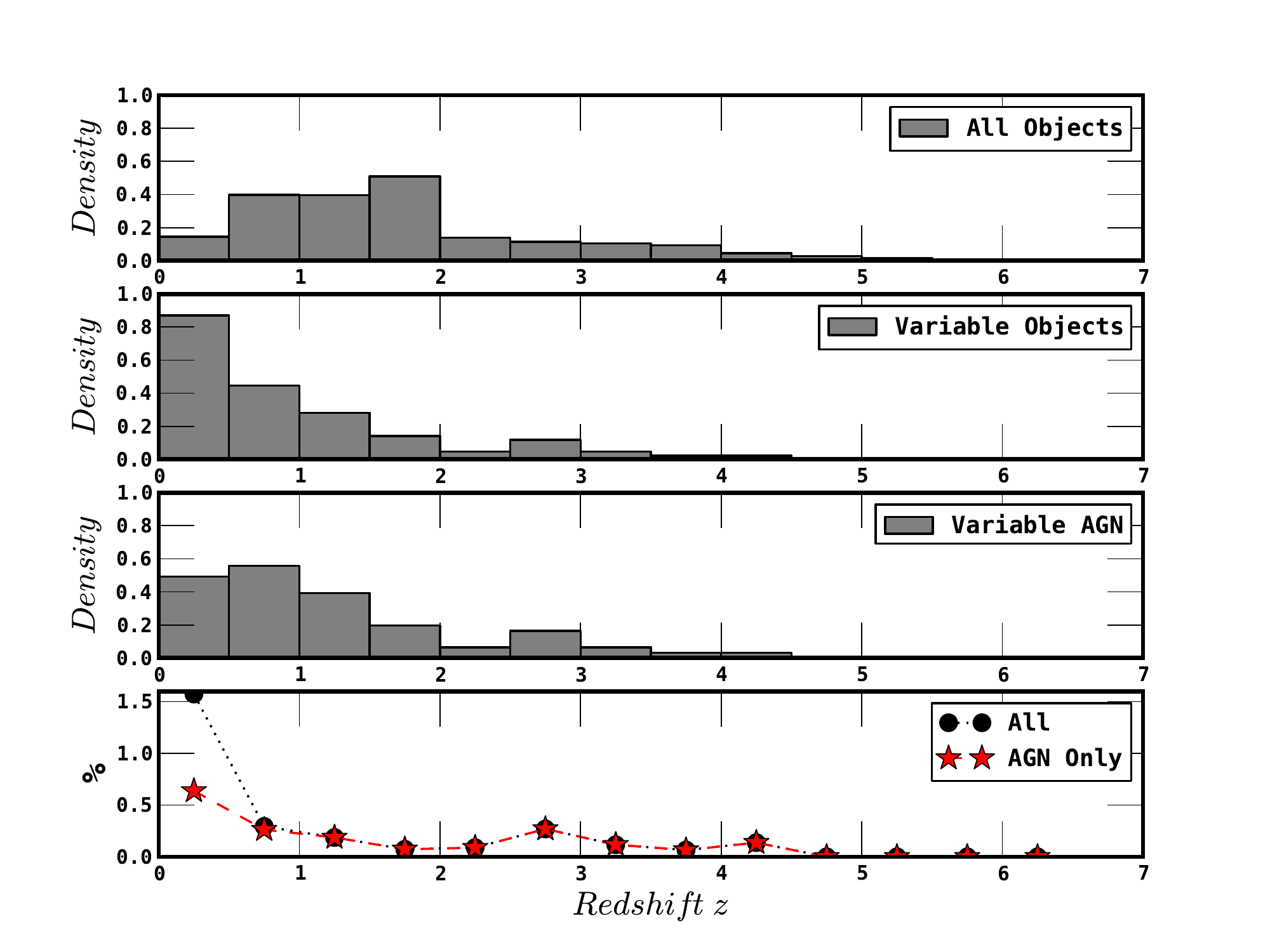}
\caption{The Distribution of photometric redshifts for all GOODS-S sources, all variability selected sources and variability selected AGN candidates only.}
\label{F:redshift}
\end{figure}

Comparing the photometric redshifts of all GOODS South sources to the AGN candidates, we see that the variability selected sources are predominantly at rather low redshift. About 1\% of all sources in the $z$ band catalogue are detected to be variable at a redshift below 0.5. The fraction of variability selected AGN candidates drops off at higher redshift. The lower signal-to-noise for objects at higher redshift makes it harder to detect variability for those sources. On the other hand, the fact that the spectrum is redshifted and therefore shorter wavelengths are sampled for higher redshift objects makes detection at higher redshifts more efficient since variability is stronger at shorter wavelengths \citep{vanden_berk_ensemble_2004,trevese_variability-selected_2008}. Our results show that the drop in signal-to-noise dominates over the favourable change in observed rest-frame wavelength (see Fig. \ref{F:redshift}). Since more stars were identified in this paper, we also show the redshift distribution with those objects removed (3rd panel from top). The overall trend does not change significantly after removing stars.

Interestingly, \cite{sarajedini_variability_2011} found an increase in the percentage of galaxies with varying nuclei with increasing redshift (their Figure 8).  The primary reason for this appears to be due to the different wavelength at which the variability surveys were conducted.  In the V-band, the distribution of galaxies is weighted towards lower redshifts when compared to the z-band survey (see Fig 8a in Sarajedini et al. as compared to the top panel in our Figure \ref{F:redshift}).  While the redshift distribution of V-band and z-band selected variables is similar, the galaxy distributions are not.  Thus, the trend in the percentage of galaxies hosting variables as a function of redshift in the two surveys differs.  The fact that the redshift distributions of the variable AGN samples are quite similar regardless of their parent populations is a testament to the robustness of using variability to identify low-luminosity AGN across a range of redshifts.

One of the main reasons to use variability was to find lower luminosity AGN at higher redshift, the absolute magnitudes of our sample are shown in Fig. \ref{F:LP_AGNMags}. As we can see, the variability selected AGN are of moderate luminosity, a majority of the sources lie below the nominal upper limit for Seyfert Galaxies of $M=-23$ \citep{veron-cetty_catalogue_2006}, showing that variability selection is capable of identifying low-luminosity  AGN. The lack of luminous AGN is not entirely surprising since those sources are generally rare in pencil beam surveys such as GOODS due to their low number density.

\begin{figure}
\includegraphics[width=8cm]{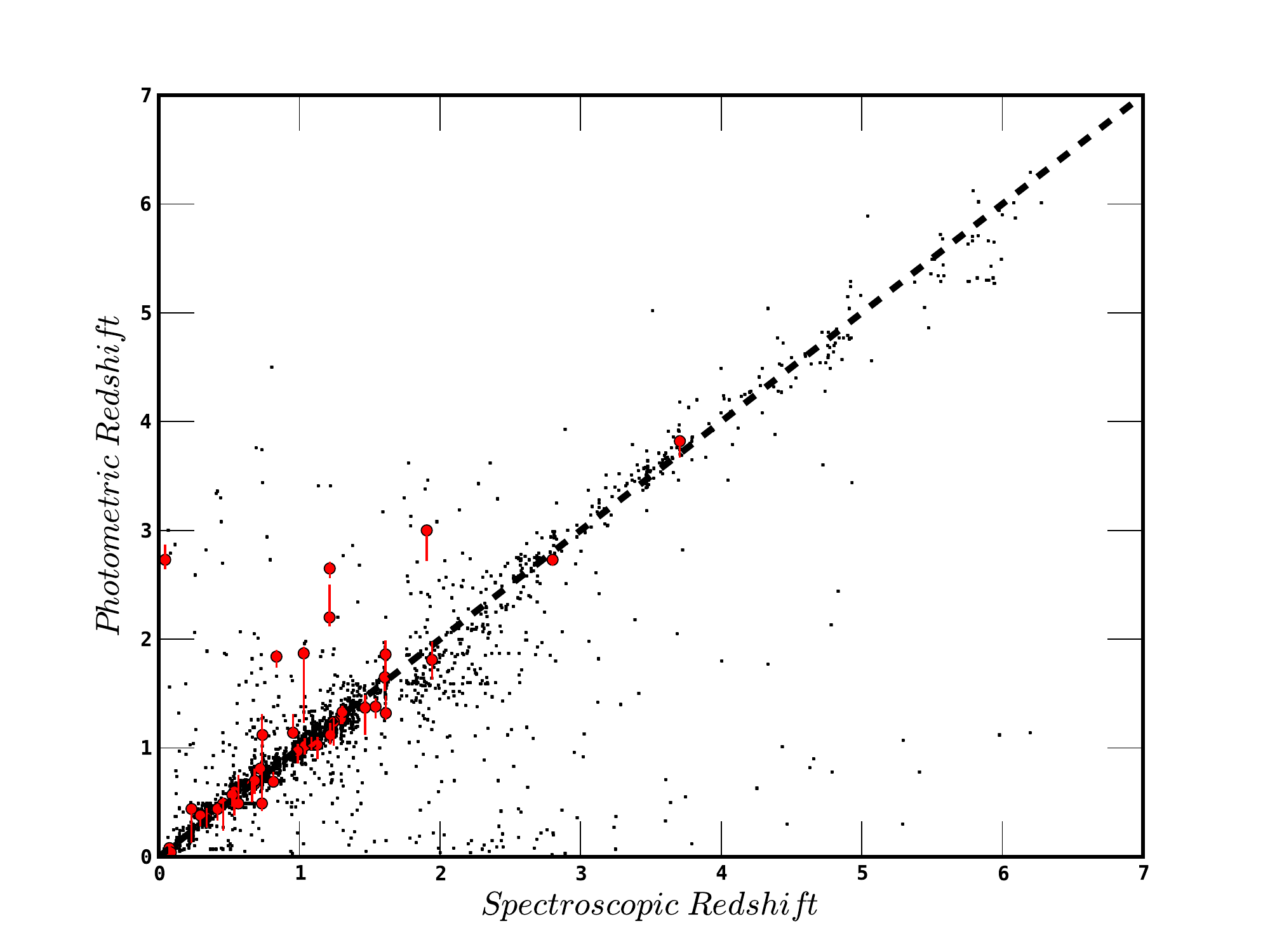}
\caption{The relation between photometric and spectroscopic redshifts from \citet{dahlen_detailed_2010}. Small black dots show all objects from \citet{dahlen_detailed_2010} and red circles show all variability selected AGN with spectroscopic redshifts, errorbars for variability selected AGN show the 68\% confidence limits.}
\label{F:z_accuracy}
\end{figure}

In photometric redshift codes, galaxy templates are fitted to multi-waveband data. Since AGN spectra are distinct from galaxy spectra, there is concern that the code might be less accurate in determining redshifts. Fig. \ref{F:z_accuracy} shows a comparison between the photometric and spectroscopic redshift for the 42 objects with spectroscopic data. We see that in most cases the agreement is very good with a few strong outliers, all of which are later on found to have SED dominated by the AGN emission. Therefore, we caution that this finding should not be generalized to the general accuracy of photometric redshifts for AGN since the AGN in this sample are generally of low luminosity.

Inspecting the SEDs of the AGN candidates shows why this is the case: the stellar IR bump stays prominent in most lower luminosity AGN, leaving a Balmer break that allows to constrain the redshift. This is explored in more detail in the following sections . On the other hand, AGN dominated sources have SEDs that make it difficult or even impossible to derive redshifts. See for example the SED of the likely blazar in Figure \ref{F:Ind_Blazar}.

\subsection{Detailed Description of Variability Selected AGN SEDs}
\label{S:subsamples}

In this section, we will discuss the SEDs of variability selected AGN in detail and argue that there are clear signs of AGN activity in a vast majority of the variability selected AGN sample.

\subsection{Properties of X-ray detected AGN}
\label{S:xray}

\begin{figure}
\includegraphics[width=8cm]{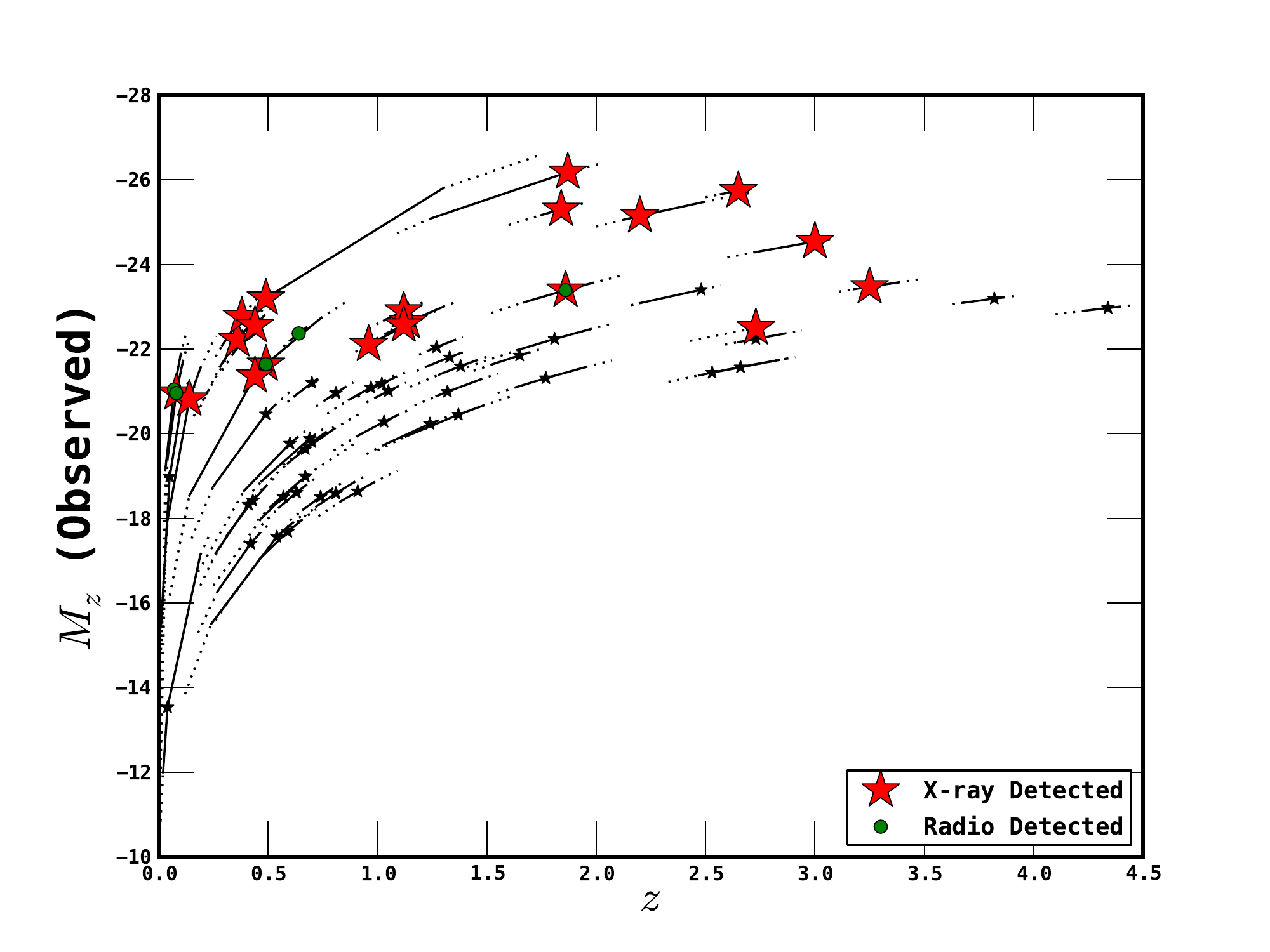}
\caption{The redshifts and observed frame $z$ magnitudes of all variability selected AGN, X-ray ands radio detections are overplotted. The magnitudes are shown for the entire object, i.e. contribution from the host galaxy is not subtracted off. The solid and dotted lines show the 68\%/95\% confidence regions for the objects with photometric redshifts.}
\label{F:LP}
\end{figure}

Next, we analyse the X-ray properties of the AGN candidates. Twenty variability selected AGN have X-ray detections, 14 were already detected in the previous 2Ms data \citep{luo_chandra_2008}. Thus six are only detected in the deeper 4Ms data.. The X-ray detected sources have a wide range  of luminosities and range from moderate to high redshifts, see Fig. \ref{F:LP}. It should also be noted that Fig. \ref{F:LP} (as well as Figure \ref{F:LP_detail}) show the integrated observed $z$ band magnitudes of the sources, i.e. the galaxy contribution is not subtracted in these plots, and actual AGN magnitudes are fainter. Additionally, since we show the same filter for the whole redshift range, the restframe wavelengths covered are quite different.

\begin{figure}
\includegraphics[width=8cm]{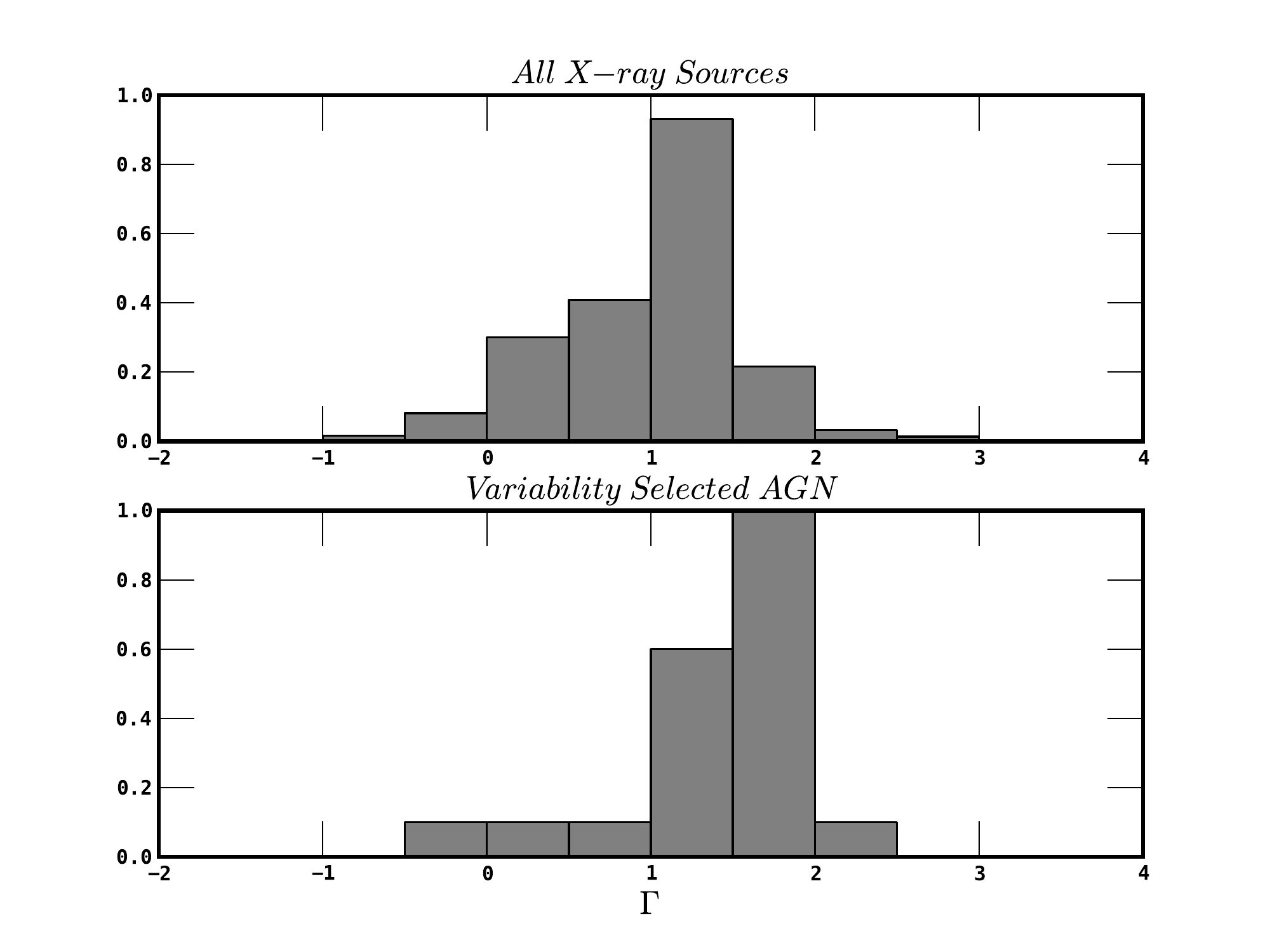}
\caption{The Distribution of $\Gamma$ for the full X-ray sample as well as the variability selected AGN (bottom) panel.}
\label{F:Gamma}
\end{figure}

Fig. \ref{F:Gamma} shows the distribution of spectral indexes $\Gamma$ from \citet{xue_chandra_2011} for the full 4 Ms sample as well as the variability selected AGN. The spectral indexes of the variability selected AGN are soft, ranging from -0.1 to 2.3. Typically the line between soft (unobscred) and hard (obscured) AGN is drawn at a $\Gamma = 1$, with larger $\Gamma$ values indicating softer spectra \citep{xue_chandra_2011}. Of our sample, only 3 sources lie below this line (J033217.06-274921.9, J033224.54-274010.4, J033228.30-274403.6). Of those three sources, two have low signal-to-noise ratios, making the derived spectral indices unreliable (J033224.54-274010.4 and J033228.30-274403.6). The other source (J033217.06-274921.9) appears to be obscured in the X-ray, it is unclear if the variability selection is due to some weak variability from obscured emission, or if the source shows a peculiar X-ray spectrum. Besides those sources, our sample includes a majority of  AGN with soft spectral indexes, as expected from the fact that variability selection strongly favours unobscured sources. This is consistent with the result of \citet{sarajedini_variability_2011} where V-band selected variables were found to have generally softer X-ray spectra based on hardness ratios.

\begin{figure}
\includegraphics[width=8cm]{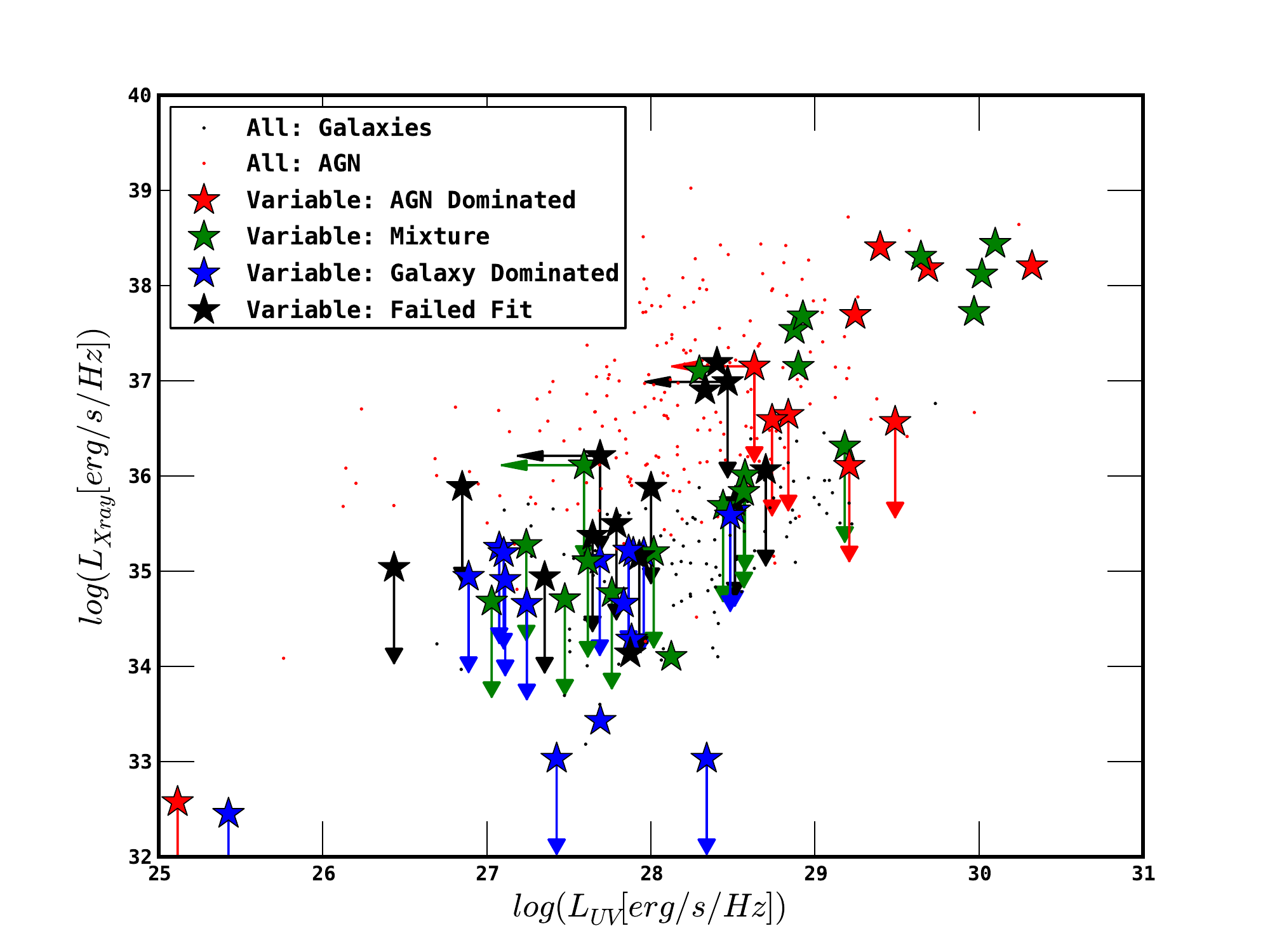}
\caption{X-ray versus UV absolute luminosity for all X-ray sources (black dots) as well as X-ray detected variability selected AGN. Xray fluxes are full Chandra fluxes (0.5-8kev).}
\label{F:alphaox}
\end{figure}

The optical to X-ray ratio of AGN has been studied in detail and shows a rather small dispersion \citep{brandt_nature_2000,strateva_color_2001}. Fig. \ref{F:alphaox} shows the correlation between the 0.5-8kev X-ray flux and optical
luminosity for both the AGN candidates and the general 4Ms catalogue, for sources without X-ray detection, we show the catalogue upper limits. The AGN candidates follow the general trend of the X-ray catalogues, however, they avoid the area of very X-ray dominated AGN. The AGN candidates cover almost the whole range in X-ray luminosities. X-ray detections are provided in Table \ref{T:AGN} in the Appendix.

Nine of the twenty x-ray detected AGN are morphologically dominated by the AGN. Eleven are clearly extended. Of the nine point source dominated objects, only two have clearly visible host galaxies. The absorption corrected X-ray luminosities for the point sources with X-ray detection range between $2.3 \times 10^{43} - 3.6 \times 10^{44}$ erg/s placing them just below the knee of the X-ray luminosity function \citep{aird_evolution_2010}.

The eleven extended sources with X-ray detections have luminosities lower than those of the point sources, ranging between $3.4\times 10^{39} - 4.6 \times 10^{43}$ erg/s, ranging all the way from low luminosity to moderatly luminous AGN. This reaches down to the limits found for local ultra low luminosity AGN \citep{ho_nuclear_2008}. The morphologies of the extended X-ray detected AGN are varied: three late-type galaxies, three early-type galaxies, three disturbed systems as well as two clear disk-bulge mixtures. Two of the disks show clear central point sources, one shows a weak tidal tail, one of the irregular systems is classified as a train-wreck merger.

Out of the 20 X-ray detected sources, five show SEDs clearly dominated by the galaxy, with best fit SEDs having very low levels of star formation. Inspection of the ACS B-band imaging data for all objects with galaxy dominated SEDs show clear central compact sources.

\begin{figure}
\includegraphics[width=8cm]{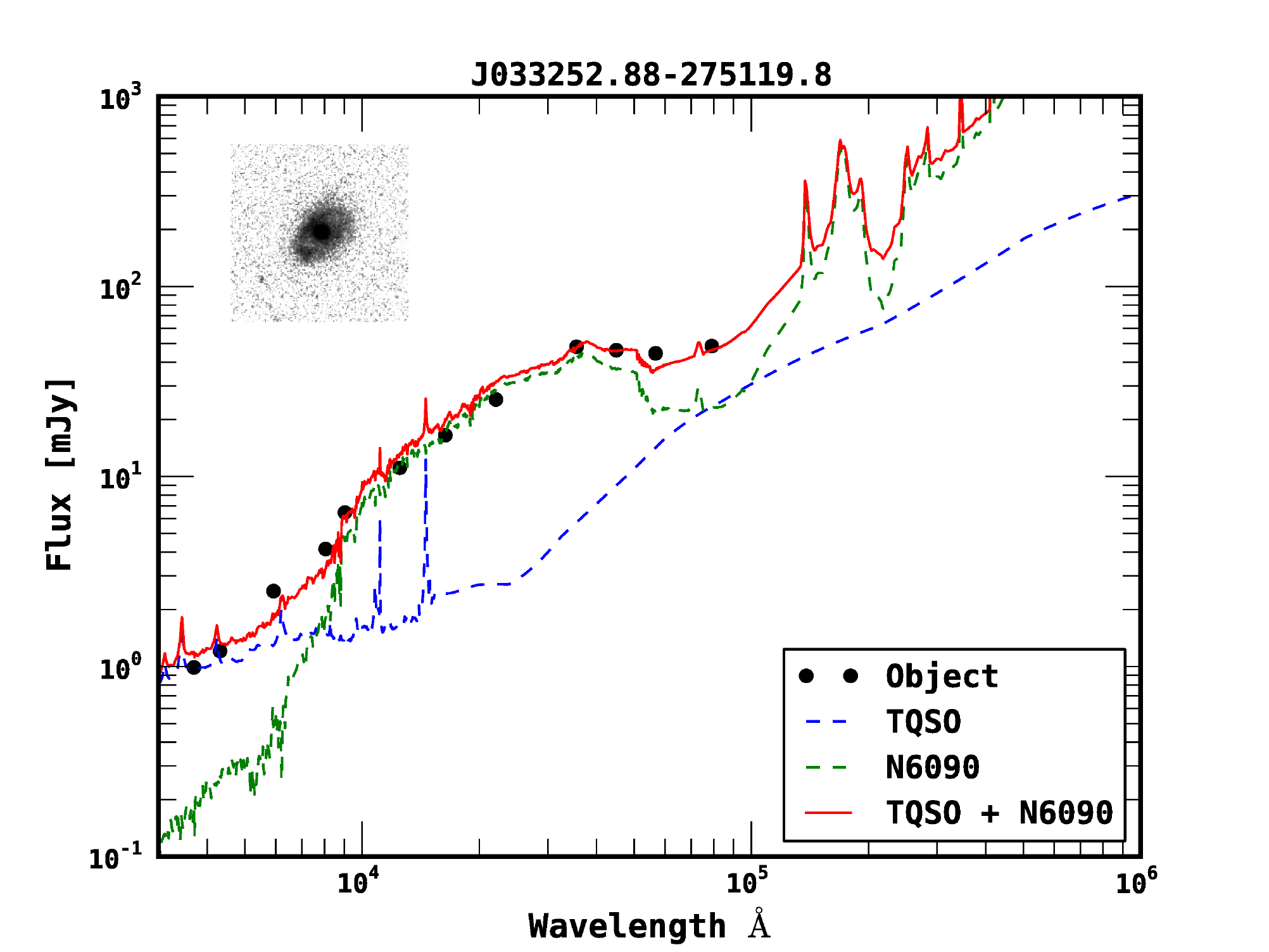}
\caption{Multiwavelength SED of J033252.88-275119.8, overlaid the best fit mixture. Inset plots shows a $5\times5$ arcsec cut-out of the object in $z$.}
\label{F:Ind_XrayNice}
\end{figure}

Seven objects are classified as mixture objects, with AGN contributions in $z$ ranging from 20-50\%. All of these show clear excess both bluewards of the Balmer break and in the IRAC bands with respect to galaxy templates. A representative example is shown in Fig. \ref{F:Ind_XrayNice}.

The five objects with quasar dominated SEDs are all point sources. Their SEDs are relatively flat spectra over a wide wavelength range. Redshifts for those objects are problematic. Several redshifts seem wrong even though they are spectroscopic. The featureless spectra make it difficult to determine proper redshifts.

\begin{figure}
\includegraphics[width=8cm]{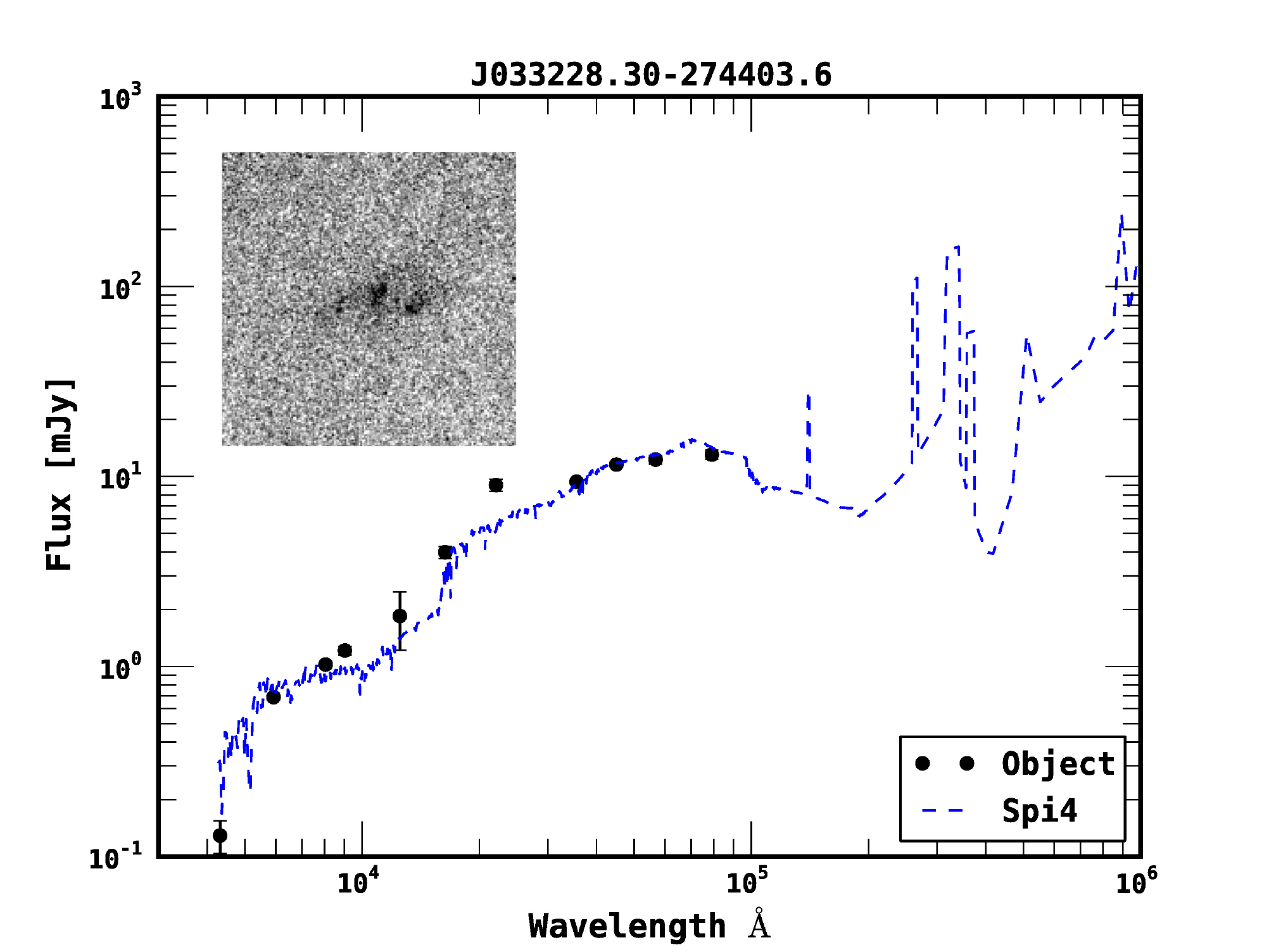}
\caption{Multiwavelength SED of J033228.30-274403.6, overlaid the two best fit SED. Inset plots shows a $5\times5$ arcsec cut-out of the object in $z$.}
\label{F:Ind_AwesomeSeyfert}
\end{figure}

One spectacular variability selected object with X-ray detection is J033228.30-274403.6. The object has a photometric redshift of 3.25. The SED is fit reasonably by either a Spiral or a Seyfert 1.8 SED. However, there are erratic jumps in the SED that indicate variability. The object is detected in the soft X-rays, but not the hard X-rays. The X-ray luminosity is $2.12\times10^{43}$ erg/s. The SED, as well as the image in $z$ is shown in Fig. \ref{F:Ind_AwesomeSeyfert}. The objects is relatively faint, extended and highly disturbed. This fascinating source shows the incredible power of variability selection.

\subsection{Properties of radio detected AGN}
\label{S:radio}

Radio emission is prominent in only about 10\% of all AGN, however, in those radio-loud objects it can be significant. We therefore investigate the radio properties of the AGN candidates. Only five sources have radio detections. Of the five radio-detected variable AGN, three are radio loud, J033210.91-274414.9, J033217.14-274303.3 and J033239.47-275300.5. J033210.91-274414.9 and J033217.14-274303.3 are also detected in the X-ray. Morphologically, they are distinct, J033210.91-274414.9 and J033217.14-274303.3 are point sources, while J033239.47-275300.5 is located in a close pair of galaxies with a weak stellar bridge. Of the five radio detected AGN, two are low-redshift spirals and three are point-like objects. Three of the radio objects are also detected in the X-ray, two are compact sources and one is a low-redshift spiral galaxy. The SEDs of the radio detected sources are discussed in more detail below. As expected, radio selection only identifies a small subset of AGN. Radio detection are provided in Table \ref{T:AGN} in the Appendix.

\begin{figure}
\includegraphics[width=8cm]{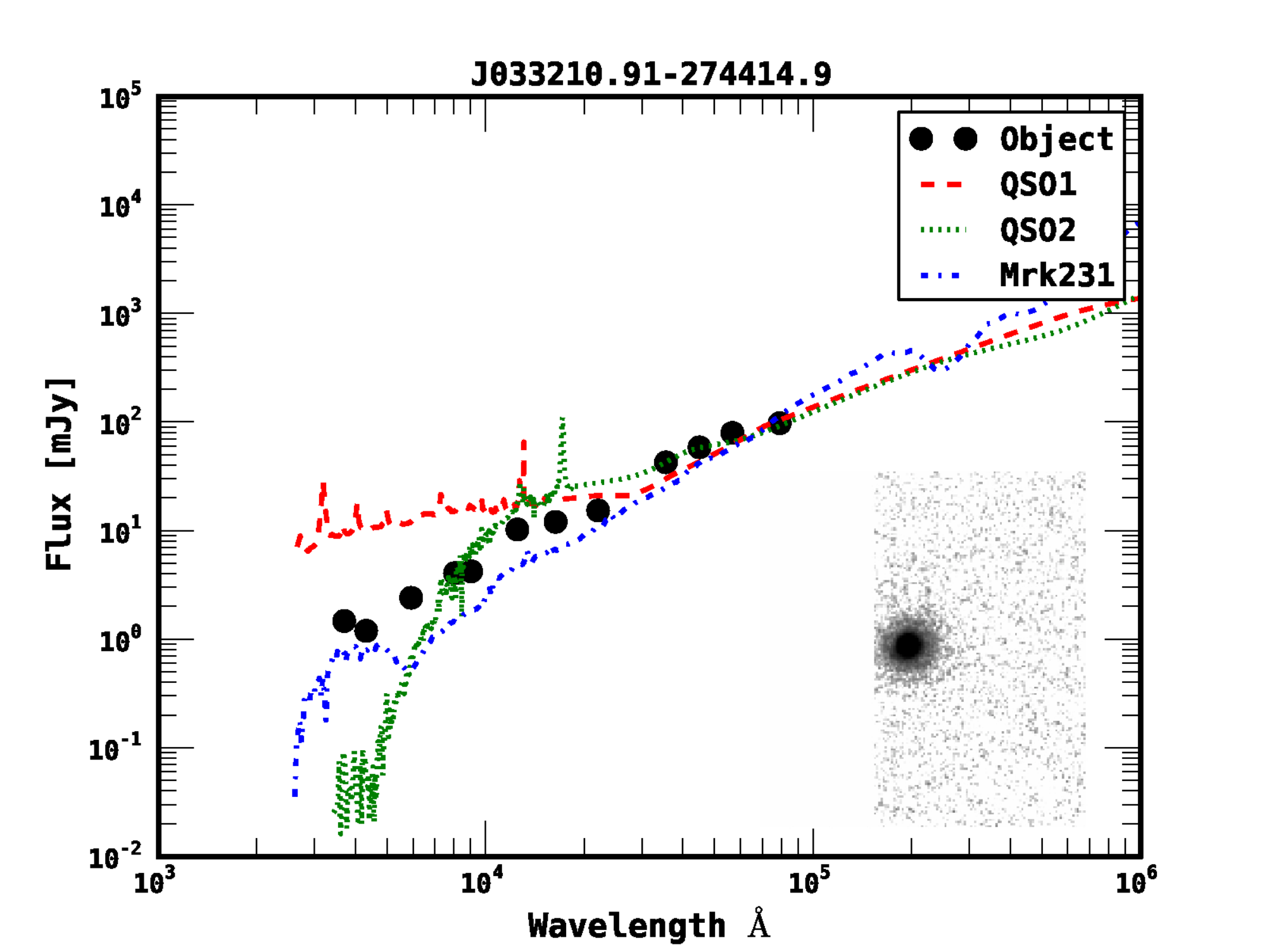}
\caption{Multiwavelength SED of J033210.91-274414.9, overlaid several of the best fitting SEDs, redshifted to the spectroscopic redshift of 1.613 derived for this object. The best fitting SED was QSO1, followed by QSO2 and Mrk231. Inset plots shows a $5\times5$ arcsec cut-out of the object in $z$.}
\label{F:Ind_Blazar}
\end{figure}

J033210.91-274414.9 has a spectroscopic redshift of 1.6 and a photometric redshift of 1.86. It is extremely radio-loud, with a value $log(R)\sim3.5$, it is also detected in the X-rays with $\Gamma=1.6$ and has an absorption corrected absolute luminosity of $2.38\times10^{44}$ erg/s, making it one of the most luminous objects in this sample. The object has an absolute magnitude in $z$ of -23.04. The entire SED is shown in Fig. \ref{F:Ind_Blazar}. The SED is rather peculiar. The optical-NIR spectrum is a power-law. Such shapes are usually common in blazars: highly beamed, extremely variable AGN \citep{urry_unified_1995}. The spectra of this objects shows broad, but relatively weak, emission lines. The radio spectrum has a slope of $\alpha=0.5$, putting it just on the border between Flat Spectrum Radio Quasars (FSRQs) and Steep Spectrum Radio Quasars (SSRQs). This suggests that the objects is likely a blazar.

J033217.14-274303.3 has a spectroscopic redshift of 0.566 and a photometric redshift of 0.49. It has a radio-loudness $log(R)\sim1.7$ and is also detected in the X-ray ($\Gamma=1.87$). The radio spectrum is rather flat ($\alpha=0.5$). An absolute magnitude in $z$ of -22 puts it below the limit for quasars. The object is clearly extended with a early-type morphology. The object shows a clear contribution from a stellar bump as well as a steep rise at 8$\mu m$, typical of many of the variability selected objects discussed in the following sections.

J033239.47-275300.5 does not have a spectroscopic redshift. The photometric redshift is 0.64. It is radio-loud with $log(R)\sim2.2$ and is not detected in the X-rays. With an $\alpha$ of 0.4, it is the radio-detected object with the flattest spectrum in our sample. The AGN is located in an elliptical galaxy that is located in a galaxy pair. The second galaxy is also spheroidal but more symmetric than the galaxy with the AGN. They seem to be connected by a faint stream, suggesting that they are actually undergoing a merger. The
projected distance between the two galaxies is about 1.2 arcsec, which corresponds to approximately 8kpc at the redshift of the object. The SED is galaxy dominated. Apart from the extremely strong radio flux, there is no clear sign of AGN activity, however, we decide to keep this in the sample to due its radio-loudness.

Two of our objects (J033229.88-274424.4 and J033229.99-274404.8 ) are clearly extended spiral galaxies at very low redshift (both objects have a spectroscopic redshift of 0.076). They show clear extended spiral galaxy patterns as well as clear central point sources. Both objects have radii of about 2 arcseconds. Both objects are detected in radio but are radio-quiet ($R\sim2-3$). Given the fact that the AGN contribution to the overall flux in these objects is lower than 10\%, this is a clear lower limit and the AGN itself is radio-loud in both objects. Both objects are very bright, with absolute $z$ magnitudes around 16.5, making them moderately bright spiral galaxies with $z$ band magnitudes around -21. One of the objects has spectroscopic data and shows a relatively featureless spectrum with only weak lines.

Due to the small contribution from the central point source, the X-ray emission is so weak that of those two object, only one (J033229.99-274404.8) is detected in the X-rays, with a low X-ray luminosity of only $1.3\times 10^{40}$ erg/s and $\Gamma=1.6$. The other object remains undetected, even in the 4Ms data. Interestingly, the two radio-detected spirals are at the same redshift and are only separated by a projected distance of 30kpc. It is unclear if they are gravitationally bound or if tidal forces might have contributed to AGN activity in both of those sources, this might fit with an emerging picture that close interaction without actual merging enhance AGN activity \citep[e.g.][]{ellison_galaxy_2011,koss_understanding_2012}.

\subsection{SEDs of AGN candidates not detected in X-ray or radio}

In the following sections, we will shortly discuss those sources not detected in X-ray or radio, we will start by identifying false positives and then continue on to different subsamples of AGN, divided by percentage AGN contribution.

\subsubsection{Likely False Positives}
\label{S:contaminants}

While a large majority of objects in the sample show clear signs of AGN activity in their SEDs, in a few cases, the SEDs are very well described by a galaxy SED without other signs of AGN activity. In particular, there are six cases which might possibly not be AGN but false positive detections, we will shortly discuss their properties.

J033241.87-274651.1 is a 'tadpole' type galaxy \cite[see e.g.][for a description of this class of galaxies]{straughn_tracing_2006}. The best fit SED is a Spiral galaxy SED, but the actual SED appears slightly reddened. While weak AGN activity in the 'head' is not excluded, this object is probably a false positive.

J033219.86-274110.0 is entirely galaxy dominated, with a best-fit 'Sd' SED. The rest-frame UV imaging data shows no clear central point source. The object appears to be undergoing a merger. It is very likely a false positive.

J033210.52-274628.9 appears to be somewhat reddened with respect to a normal S0 SED, there is no clear evidence of AGN activity in the SED. The object is very compact but resolved. It is likely a false positive.

J033246.37-274912.8 is reasonably well fit by a S0 SED, the morphology shows a clear mixture of bulge and disk components, there is no central point source visible in the rest-frame UV frames, this is likely a false positive.

J033209.57-274634.9 is reasonably well fit by a Sdm SED. There is a red excess that can be fit by including a QSO component, however, there is no UV excess. This might be explained by obscuration depressing the UV excess from star formation or possibly by invoking an UV-weak AGN. This source is possibly a false positive.

J033225.10-274403.2 is well fit by an S0 SED, it shows a clear disk+bulge morphology. It is at a redshift of 0.076 and very clearly extended. The UV excess in this source is extremely weak. This is likely a false positive. 

We do caution that  while those cases are likely false positives, these objects are marked in Table \ref{T:AGN} in the Appendix. Note that all of these sources are in the 'normal' catalogue of variability selected sources \citep{villforth_new_2010} that is expected to contain around six false positives in the GOODS-S field. Therefore this false positive detection rate is expected. We no longer consider false positives in the following discussion.

\subsubsection{Quasars}
\label{S:quasars}

A total of ten sources have quasar dominated SEDs, four of which are X-ray detected, one source is detected in both radio and X-ray, leaving six sources with quasar dominated SEDs but lacking X-ray and radio detection. Redshifts for these sources are relatively unsecure.

\subsubsection{Seyferts}
\label{S:seyferts}

\begin{figure}
\includegraphics[width=8cm]{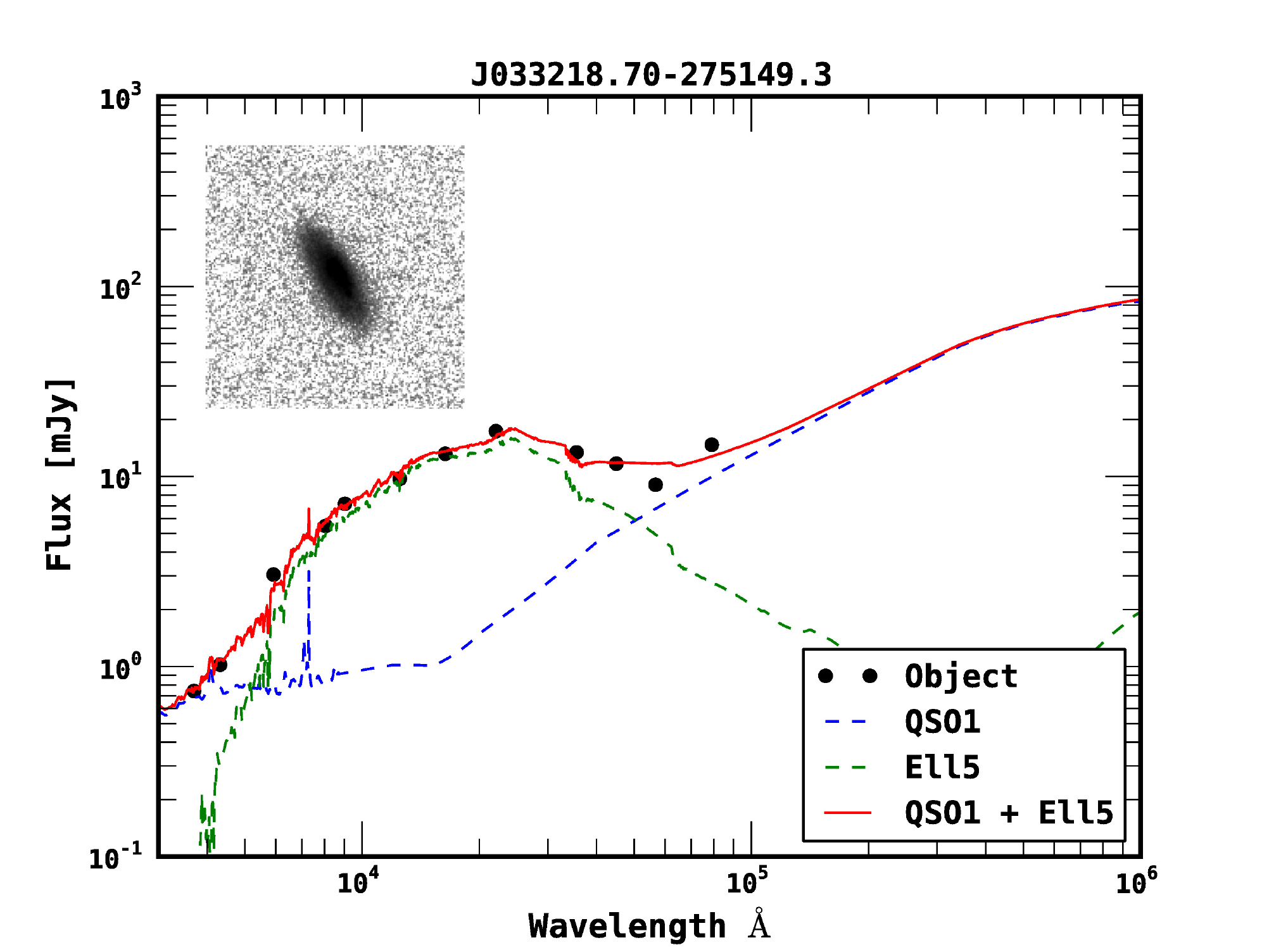}
\caption{Multi-wavelength SED of the Seyfert type object J033218.70-275149.3, overlaid the best fit mixture model. Inset plots shows a $5\times5$ arcsec cut-out of the object in $z$.}
\label{F:Ind_Seyfert}
\end{figure}

As Seyferts, we label objects not detected in the X-rays but with SEDs classified as mixtures (the AGN contributes between 10-90\% in the $z$ band). A total of 19 objects show mixture fits, of which nine are detected in neither X-ray nor radio. A representative example of this group is shown in Figure \ref{F:Ind_Seyfert}. Similar to X-ray detected AGN with mixture SEDs discussed in the previous Section, these sources show clear excesses both bluewards of the Balmer break as well as in the IR.

\subsubsection{Ultra-low luminosity AGN}
\label{S:faint_AGN}

\begin{figure}
\includegraphics[width=8cm]{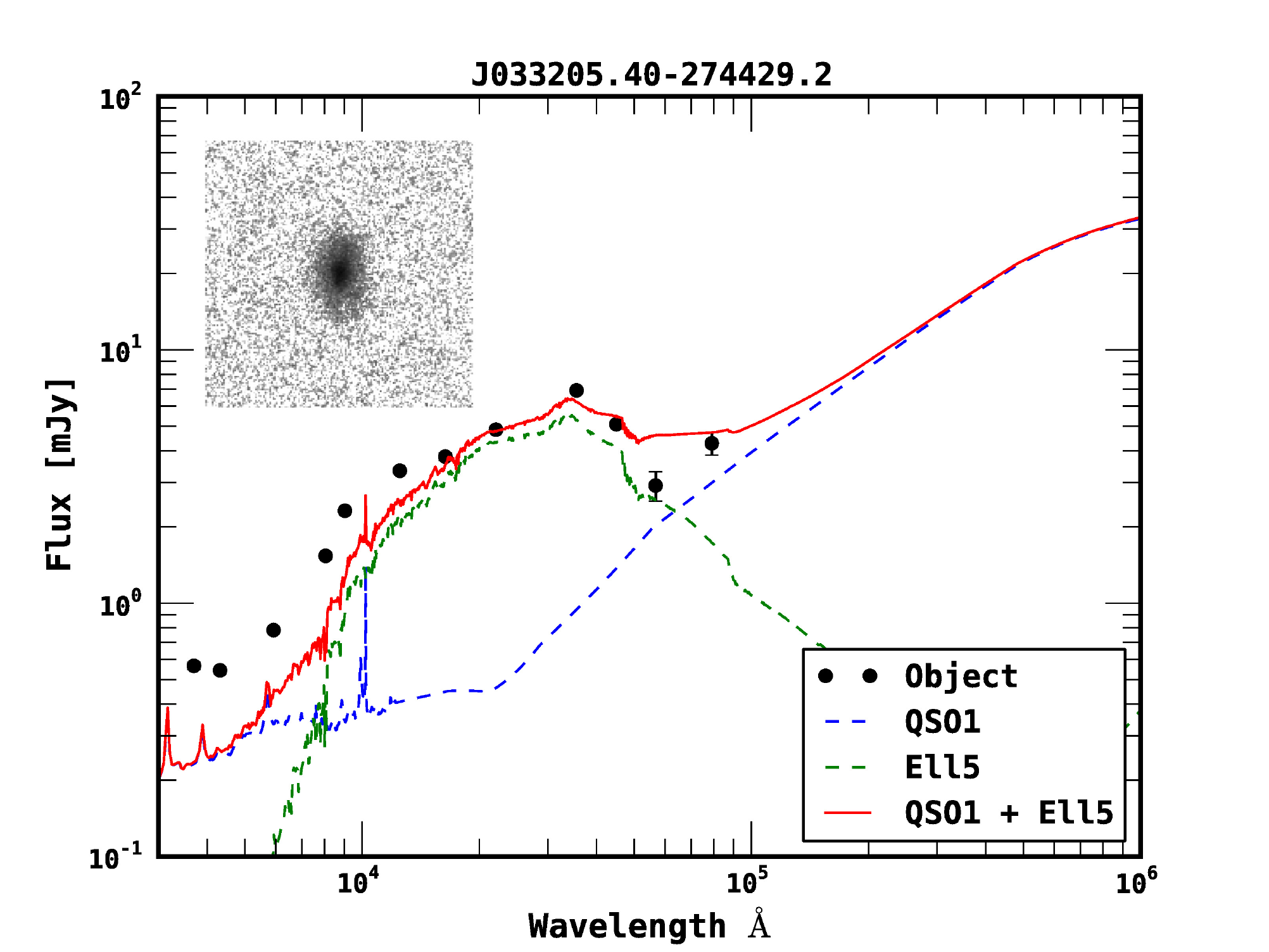}
\caption{The Ultra-low luminosity AGN J033205.40-274429.2, overlaid the best fit mixture model. Inset plots shows a $5\times5$ arcsec cut-out of the object in $z$.}
\label{F:Ind_ULLAGN}
\end{figure}

As ultr-low-luminosity AGN, we label those sources for which the mixture fits indicate AGN contributions below 10\%. 14 objects belong to this class, of which eight lack both X-ray and radio detection. In general, their SEDs resemble the SEDs of Seyfert type objects, showing excesses in both the UV and MIR. However, for those objects the excesses are notably weaker. The extra component needed to fit the SED is in general bluer than the QSO template SED, indicating a hotter accretion disk. This is expected for less massive black holes \citep[e.g.][]{laor_massive_1989}. A representative example is shown in Fig. \ref{F:Ind_ULLAGN}.

\subsection{Failed Fits}
\label{S:failed_fits}

\begin{figure}
\includegraphics[width=8cm]{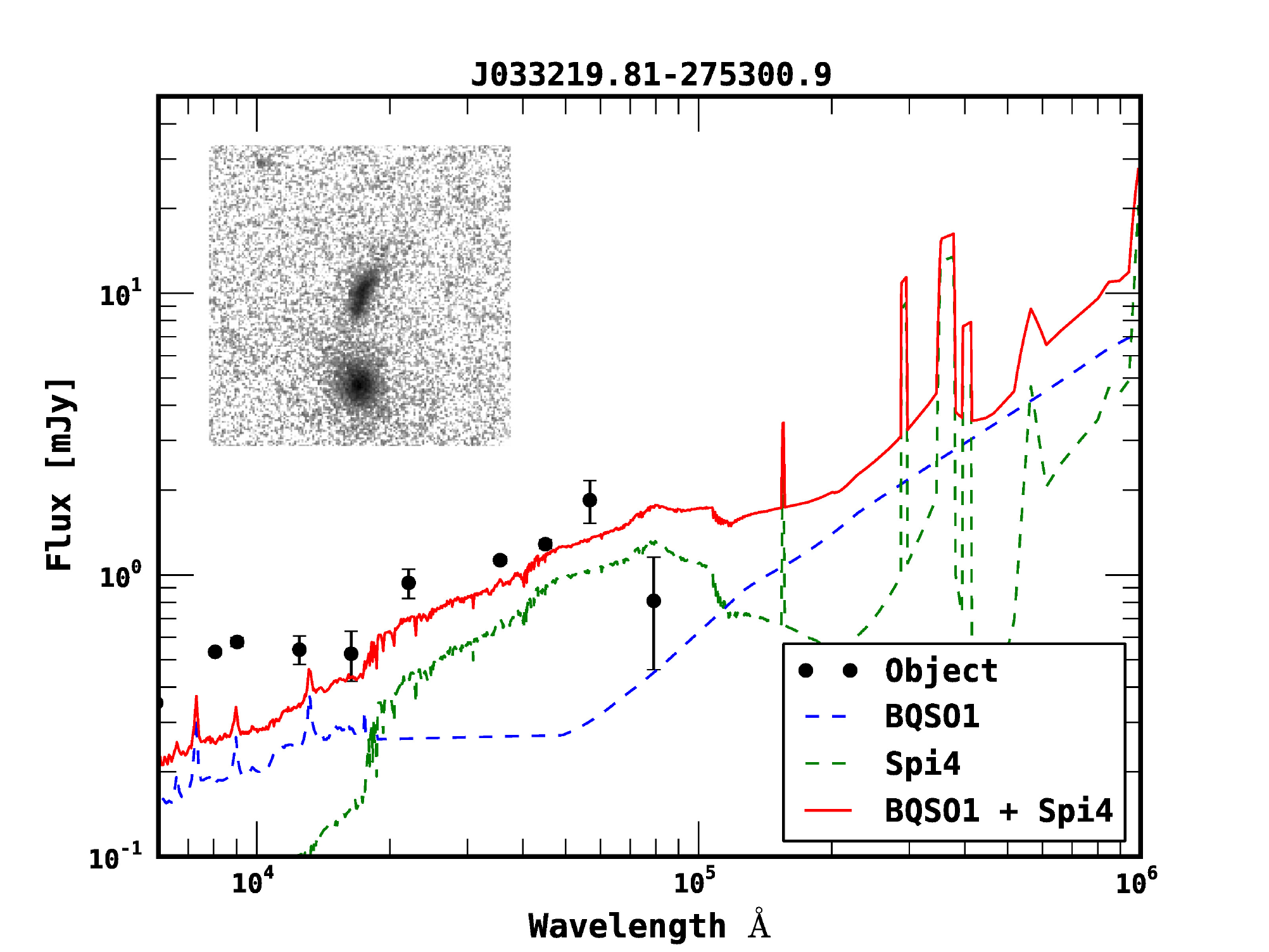}
\caption{SED of J033219.81-275300.9 with best fit mixture SED overlaid, the SED is clearly bumpy ad yields a poor fit for all templates. Inset plots shows a $5\times5$ arcsec cut-out of the object in $z$.}
\label{F:Ind_bumpy1}
\end{figure}

The failed fit sources are well fit neither by a mixture model, single galaxy or AGN SED nor a stellar SED. 12 objects belong in this group. Their SEDs show erratic bumps that are indicative either of variability or strong line emission, failed fits objects show no clear SED features such as breaks that are reproduced by the templates. Figure \ref{F:Ind_bumpy1} shows a representative example of this group, apart from the poor fit, it is apparent that there is no clear stellar IR bump that it usually needed to provide a reasonable fit. Failed fit objects are generally faint but extremely compact, their featureless SEDs and erratic bumps indicate that they are highly variable and likely quasar dominated. Since no good fit is found for those objects, their redshifts are poorly constrained. Carefully modelling the variability across the wavelength regime in these sources might  help recover the underlying SED, but this is beyond the scope of the current paper.

\subsection{IRAC Colours}
\label{S:irac}

\begin{figure*}
\includegraphics[width=8cm]{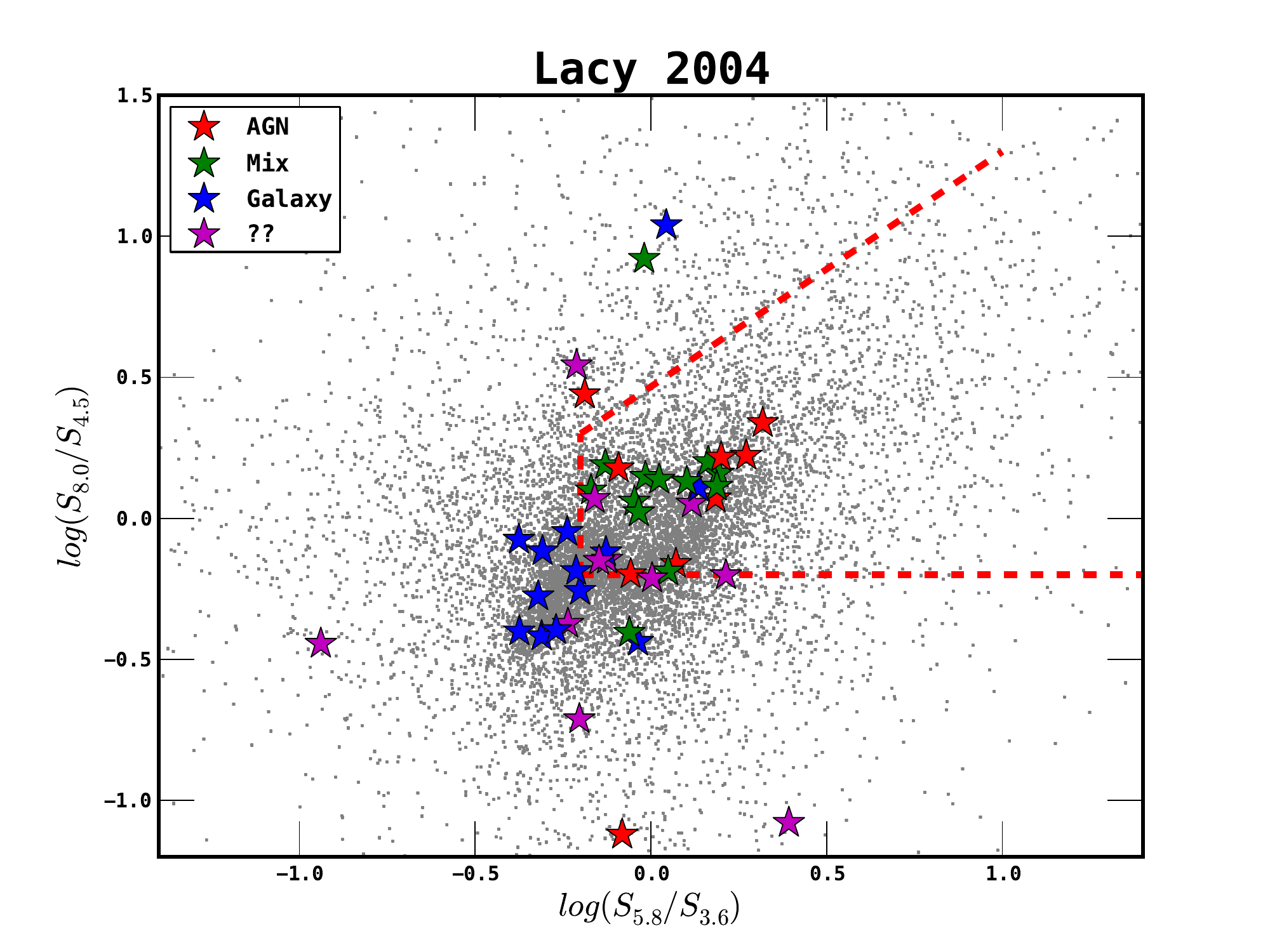}
\includegraphics[width=8cm]{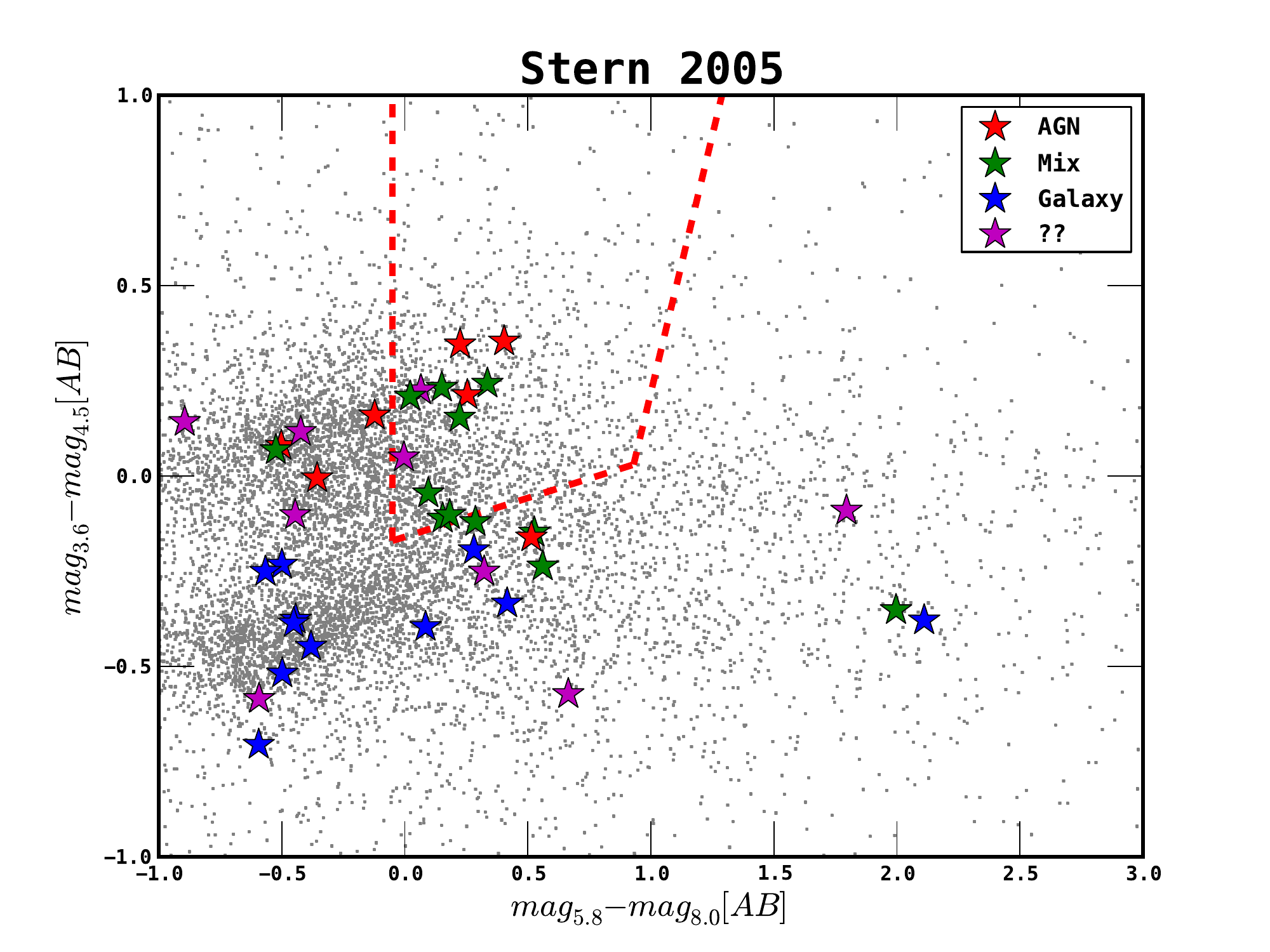}
\caption{Lacy et al. (2004) and Stern et al. (2005) color-color plot for AGN selection. The dashed lines shows the AGN 'wedge'. Small black dots are all objects from Dahlen et al. (2010) with detection in all Irac bands. Stars show variability selected AGN, color-coded by AGN contribution, based on SED fitting. False positives are not shown.}
\label{F:Irac}
\end{figure*}

IRAC colour selection has commonly been used as a way to select AGN. Due to the fact that the dust in AGN is heated to a higher temperature than in normal starforming galaxies, their IR SEDs show considerable differences \citep[e.g.][]{sanders_warm_1988,lacy_obscured_2004,stern_mid-infrared_2005,donley_spitzers_2008}. Two different IRAC colour-colour plots have been suggested to separate AGN from star-forming galaxies: \citet{lacy_obscured_2004} and \citet{stern_mid-infrared_2005}, both defined a wedge within which most objects should be AGN dominated. Figure \ref{F:Irac} shows the colour-colour plots for AGN candidates as well as the full GOODS-S Sample. For the AGN dominated sources with IRAC detection, 80\% are in the Lacy Wedge and 50\% are inside the Stern wedge. For the Mixture objects 83\%/54\% are inside the wedge for Lacy and Stern respectively. For the galaxy dominated sources, those numbers drop to 12\%/0\% for Lacy and Stern respectively. For the objects with failed fits, 54\%/ 18\% would be flagged as possible AGN using the Lacy and Stern correction respectively. This shows that both selection methods recover moderately luminous AGN but are not suitable to detect the lowest luminosity AGN.

\cite{donley_spitzers_2008} have studied the reliability of different AGN IR selection techniques and found that contamination of normal star-forming galaxies is rather common in both the Lacy and Stern wedge, especially when low-luminosity sources are included. They showed that the colour-colour selection will include most actual AGN, but with a significant contamination. However, here we demonstrate that  low-luminosity AGN can be located outside the Stern and Lacy wedges. We therefore argue that IR colour-colour selection misses most low luminosity systems in which the SED is dominated by the galaxy, as also noted in \citet{donley_identifying_2012}.

\subsection{Summary of SED Fitting Results}

Here we will short summarize the results presented in this section.

SED fist were performed on 71 of the initial 88 sources found in GOODS-S by \citet{villforth_new_2010}, the rejected 14 objects being stars previously identified in the literature. SED fits revealed another ten stars, mostly of G-M type, listed in Table \ref{T:Stars} in the Appendix but further disregarded in the analysis, this leaves 61 AGN candidates.

For the remaining 61 AGN candidates UV-IR AGN/Galaxy mixture SED fits were performed, six are likely false positives, in good agreement with the expected number of false positives in the sample. This leaves 55 confirmed AGN. 43 had successful fits, leaving 12 failed fits, discussed in Section \ref{S:failed_fits}. Twenty sources have X-ray detection, five have radio detection, three are detected in both X-ray and radio. Of the 43 sources with successful fits, ten are quasar dominated, 19 are mixture objects and 14 are galaxy dominated. Especially the low luminosity AGN would be missed by IRAC selection techniques.

In summary, we show that variability selection can reliably identify AGN far below the limits of even extremely deep X-ray data and not identified using commonly used IRAC colour-colour selection methods. SED fits where succesful for 43/61 sources, with the remaining SEDs showing clear non-stellar SEDs with clear bumps, indicating that they are indeed AGN.

\section{The host galaxies of variability selected AGN}
\label{S:hosts}

\subsection{Morphologies of variability selected AGN}
\label{S:morph}

The properties of AGN host galaxies are of importance for understanding the triggering mechanisms of AGN. In this section, we will discuss the morphologies of the host galaxies in this sample. Morphological Classification is performed by eye. The classifications scheme is described in detail in the Appendix, and all morphological classifications are provided in Table \ref{T:AGN}. In the following discussion we will distinguish between unresolved sources, disks, bulges, disk-bulge mixtures, mergers and unclassifiable.

Figure \ref{F:LP_detail} show the morphologies in the luminosity-redshift parameter space. Interaction is common, especially at the low luminosity end. At lower redshift, late-type morphologies dominate. No AGN is hosted by an early-type galaxy at redshifts higher than $\sim$1. Out of the 12 sources classified as late-type, eight show clear central compact sources, two sources show weak tidal tails indicating possible recent mergers. For five out of twelve late-type hosts, the orientation can be determined clearly, of those, four are face-on while only one is clearly edge on. While the numbers are too small to perform a detailed analysis, this might indicate that variability selection is more successful in face-on systems. This could either indicate that AGN are aligned with the host galaxy, or that absorption is stronger in edge-on systems.

Out of the mere five early-type galaxies, only two show very obvious central compact sources and one shows a close-by neighbour (J033239.47-275300.5, discussed in more detail in Section \ref{S:radio}). Five objects are classified as disk-bulge mixture, of which one is clearly edge-on. All of those spheroids were best fit with an S0 galaxy template and show AGN contribution in $z$ of $\sim 5\%$. 18 objects are classified as interacting, of those, three show clear central compact cores, four are very clear ongoing mergers and five are classified as clumpy or train-wreck. Of the fourteen objects without a clear morphological classification, two show compact cores, two have nearby neighbours and one shows a weak tidal tail. Implications of these findings will be discussed in Section \ref{S:discussion}.

\subsection{Red sequence and blue cloud: are variability selected AGN
experiencing quenching of star formation?}
\label{S:redblue}

\begin{figure}
\includegraphics[width=8cm]{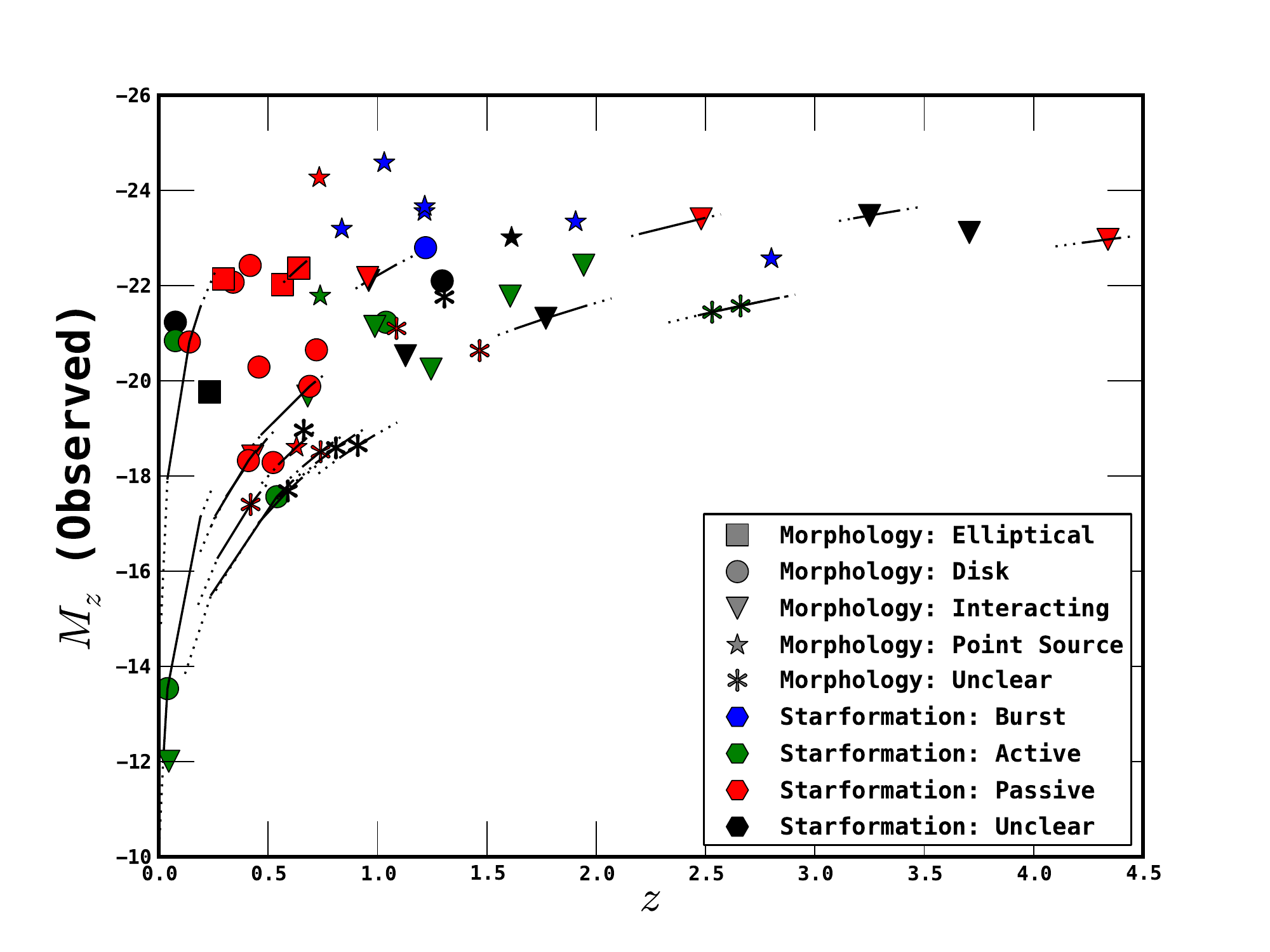}
\caption{Star formation properties and morphology of variability selected AGN in a observed frame $z$ band magnitude redshift plot. Colours indicate star formation level: red: passive, green: moderately star forming; blue: strongly star-forming. Marker style show morphology: circles: disk galaxies; down triangles: elliptical; up triangle: disturbed morphology; pentagons: disk/bulge mixtures; cross: unclear (too faint); stars: point source dominated. The magnitudes are shown for the entire object, i.e. contribution from the host galaxy is not subtracted off. False positives are not shown.}
\label{F:LP_detail}
\end{figure}

One of the questions that is commonly discussed is if and how strongly AGN activity correlates with star formation in the host galaxy. To address this question, we show the level of ongoing star formation in the host galaxies as a function of redshift and luminosity. We divide the host galaxy SEDs into three groups: purely passive (no young stellar population), mildly star forming (contribution from unobscured young star forming regions) and starbursting (young stellar population dominant, considerable reddening). The results are shown in Figure \ref{F:LP_detail}. Starburst type SEDs are common only in high luminosity AGN, low luminosity AGN hosts are dominated by either mildly star-forming or passively evolving host galaxies. The implication of these finding will be discussed in Section \ref{S:discussion}.

\begin{figure}
\includegraphics[width=8cm]{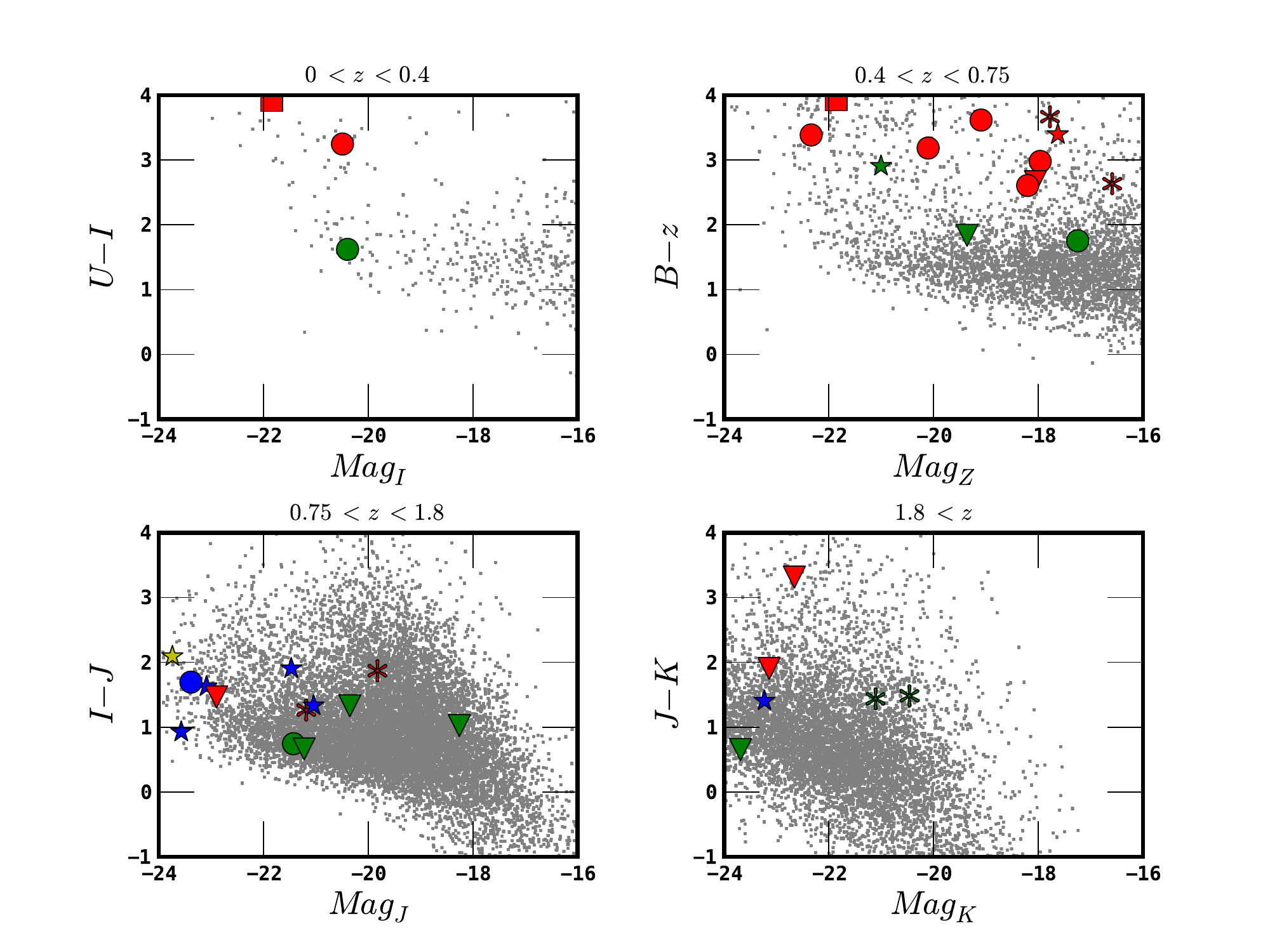}
\caption{Optical Colours of variability selected AGN host galaxies from SED Fits compared to the full sample of galaxies. Colours indicate star formation level: red: passive, green: moderately star forming; blue: strongly star-forming. Symbols indicate morphology, as described earlier in Section \ref{S:morph}: stars: point source, triangle: interacting; circles: disks; pentagons: elliptical; x: unclassified. Only sources with successful fits that have not been identified as false positives are shown.}
\label{F:redblue}
\end{figure}

It has been noticed that galaxies often fall into two rather distinct groups: massive, passively evolving 'red and dead' elliptical galaxies and star-forming, less massive spiral galaxies \citep[e.g.][]{strateva_color_2001,balogh_bimodal_2004,faber_galaxy_2007}. These groups are known as the red sequence and blue cloud. Additionaly, there is a transitional phase of green valley galaxies in which the star formation is just turning off \citep[e.g.][]{strateva_color_2001,balogh_bimodal_2004,faber_galaxy_2007}. It has been argued that the host galaxies of AGN should be located in the green valley since the AGN could be responsible for quenching the star formation \citep{silk_quasars_1998,hopkins_cosmological_2008,somerville_semi-analytic_2008}. Evidence for this scenario has been mixed \citep{cardamone_dust-corrected_2010,schawinski_galaxy_2010}. We show the location of the host galaxies of our AGN with respect to the red sequence and blue cloud in Figure \ref{F:redblue}. The AGN host galaxies are located in all areas of the colour-magnitude diagram. In the two low redshift bins, the hosts of AGN are found in the green valley as well as the red sequence. In the two higher redshift bins, the AGN host galaxies are located in the blue cloud as well as the green valley. V-band selected variable AGN hosts appear to be distributed over a range of colors as well, from the red sequence through the green valley and into the blue cloud \citep{sarajedini_variability_2011}.  A slightly higher fraction of blue cloud hosts were found among the variability selected AGN when compared to those identified in X-ray surveys.  

We caution that the technique used to determine host galaxy colours does not entirely remove the contamination from the central source and might therefore be biased. Properly accounting for this effect would require careful PSF Fitting in multiple bands which is beyond the scope of the current study. In the redshift range below $z<0.75$, the location of AGN host galaxies in the red sequence and blue cloud is comparable to the Seyfert host galaxies studies by \cite{schawinski_galaxy_2010}: the late-type hosts are predominantly located at the high-mass end and avoid the blue cloud.  

\subsection{Are more luminous AGN located in higher mass galaxies?}
\label{S:AGNvsGalMass}

\begin{figure*}
\begin{center}
\includegraphics[width=20cm]{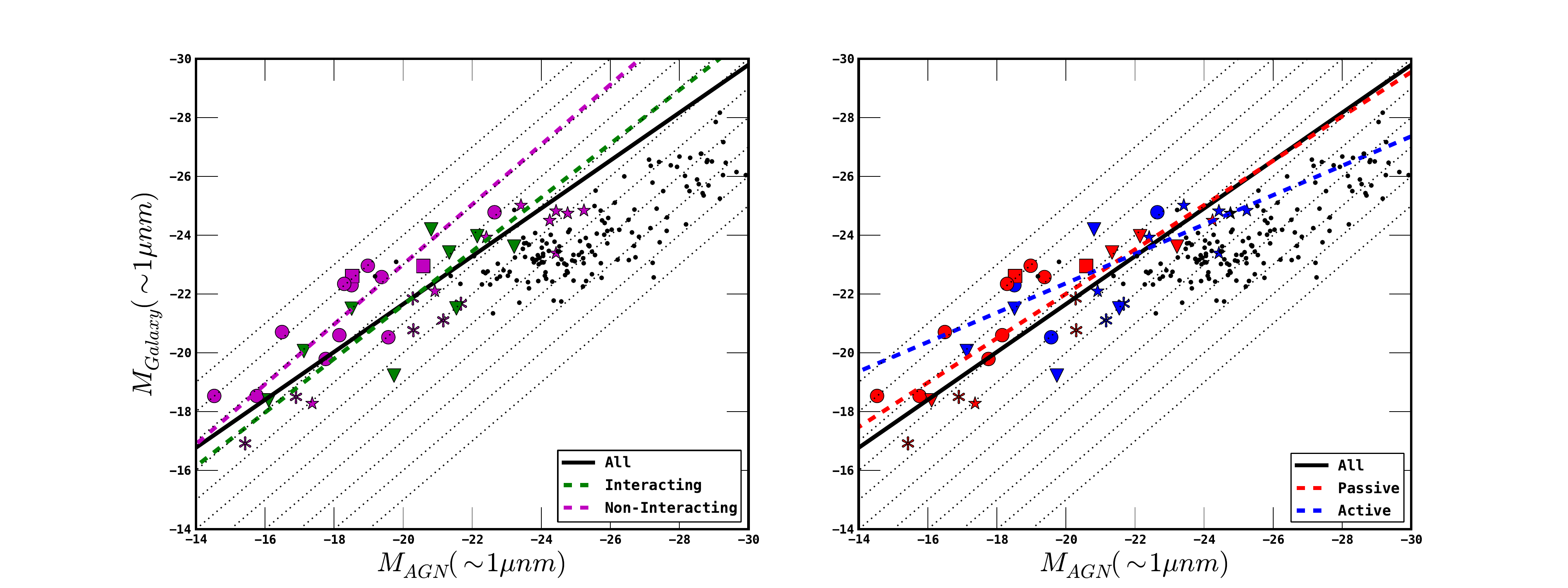}
\caption{The relation between AGN Nuclear and Galaxy magnitude at an rest-frame wavelength of $\sim 1 \mu m$. Colours indicate subsamples used for fits. In the left hand panel, those are: green: objects showing signs of interaction, magenta: objects showing no signs of interaction. In the right hand panel: red: passive, i.e. no ongoing star-formation, blue: active (ongoing star-formation). Thin black lines show constant galaxy/AGN luminosity ratios with offsets of -5 to 5 magnitudes. Symbols indicate morphology, as described earlier in Section \ref{S:morph}: stars: point source, triangle: interacting; circles: disks; pentagons: elliptical; x: unclassified. Small black dots are data from the Kotilainen et al. (2006), Kotilainen et al. (2007) and Hamilton et al. (2008). Only sources with successful fits that have not been identified as false positives are shown.}
\label{F:AGNVSGal}
\end{center}
\end{figure*}

An important aspect regarding the properties of AGN hosts is the fact that more luminous AGN seem to be hosted by more luminous and therefore more massive galaxies \citep[e.g.]{hutchings_optical_1984,bahcall_hubble_1997}. We show the relation between galaxy and AGN magnitudes for our sample in Figure \ref{F:AGNVSGal} (at restframe $\sim$ 1 $\mu$m), together with comparable data from the literature \citep{kotilainen_host_2006,kotilainen_nuclear_2007,hamilton_fundamental_2008}. Our data show that there is a clear correlation between the luminosities of AGN and their hosts. However, the luminosity of the AGN rises faster than the luminosity of the galaxy similar to findings in other studies \citep[e.g.][and references therein]{hutchings_optical_1984,bahcall_hubble_1997,schramm_host_2008}. This can be interpreted as a sign that higher mass black holes accrete at higher Eddington ratios. But it can also imply a tendency for black holes in more massive galaxies to be comparably overmassive, i.e. have higher $M_{BH}$ to $M_{galaxy}$ ratios. However, in a few cases, the galaxy luminosity is observed to rise faster at the high luminosity end \citep{schramm_host_2008}. Also, rare cases in which high luminosity AGN show no sign of host galaxy contribution are known \citep{magain_discovery_2005}.

Selection effects play an important role in this context. Systems with very bright AGN and very faint hosts (lower right corner of
Figure \ref{F:AGNVSGal}), while not generally selected against, have SEDs clearly dominated by the AGN and therefore galaxy magnitudes  are difficult to derive. On the other hand, systems with very high galaxy to AGN luminosity ratios (upper left corner of Figure \ref{F:AGNVSGal}) are selected against because the AGN in such systems are evasive and will not be identified as AGN, even using variability selection. Another effect is that high-redshift AGN are generally more luminous \citep{croom_2df-sdss_2009}, moving them towards the upper right corner. Selection effects also differ at higher redshifts: systems with high AGN-galaxy luminosity ratios are even harder to detect due to surface brightness dimming. Additionally, galaxy dominated systems are also harder to detect since low luminosity AGN are more difficult to identify at higher redshifts. This should results in a narrowing of the accessible parameter space towards higher redshift.

To analyse the correlation in more detail, we have performed a linear fit to the data for the full sample, as well as a number of subsamples. The fit for the entire sample shows a flat slope of $\sim 0.76$, similar to the slopes found in similar earlier studies \citep{hutchings_optical_1984,bahcall_hubble_1997}. This therefore indicates higher Eddington ratios in more luminous AGN or comparably overmassive black holes in these systems. The latter might indicate that those AGN are in a later stage of their black hole growth. 

Dividing the sample into subsamples according to the star formation level in the host, we find that the slope is significantly flatter for AGN with active star formation. This seems to indicate either that the tendency for higher mass black holes to show higher Eddington ratios is more strongly pronounced in actively star-forming systems, while it is weak for AGN with passive hosts. But it can also mean that star-forming systems are generally in a more advanced stage of black hole growth where the black hole is comparably more massive. This could indicate that those systems are nearing the end of their activity.

Dividing the sample into subsamples according to morphology, we find that there is no significant difference in the slope, but that there is a clear offset between the two datasets: AGN in interacting hosts having on average either higher Eddington ratios or larger $M_{BH} / M_{galaxy}$ ratios. It is however unclear if these findings are due to selection effects or if the trends found here reflect actual physical trends in the AGN population. Studying connections between host galaxies and AGN is complex, and it has been shown that carefully implementing selection effects can remove certain trends \citep[e.g.][]{aird_primus:_2011}.

\section{The environments of variability selected AGN}
\label{S:environment}

In the context of AGN triggering, it is of importance to understand the environments of AGN of different types, since this can constrain suspected triggering mechanisms such as major or minor mergers, disk instabilities or inflow of gas.

\begin{figure*}
\includegraphics[width=8cm]{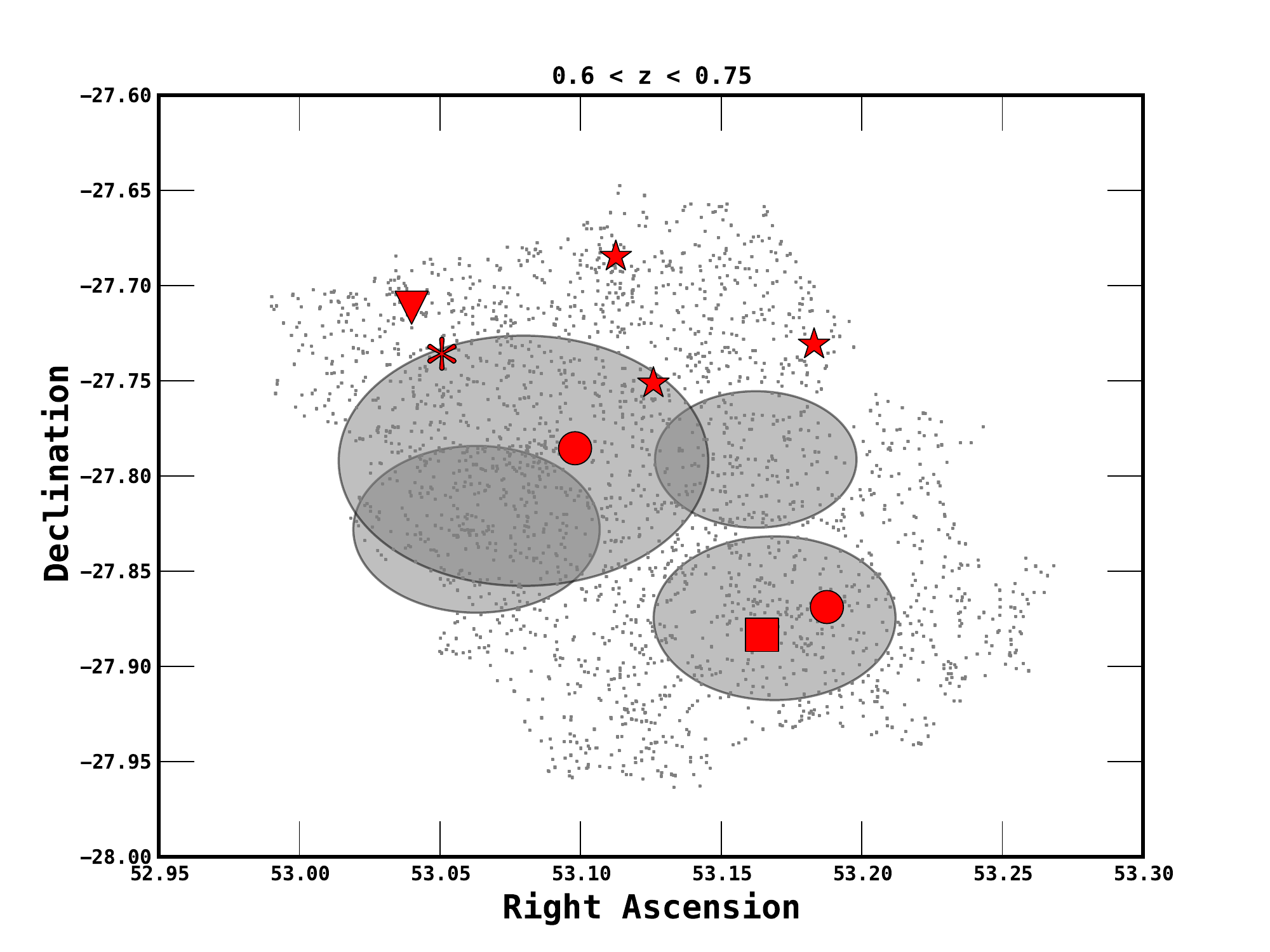}
\includegraphics[width=8cm]{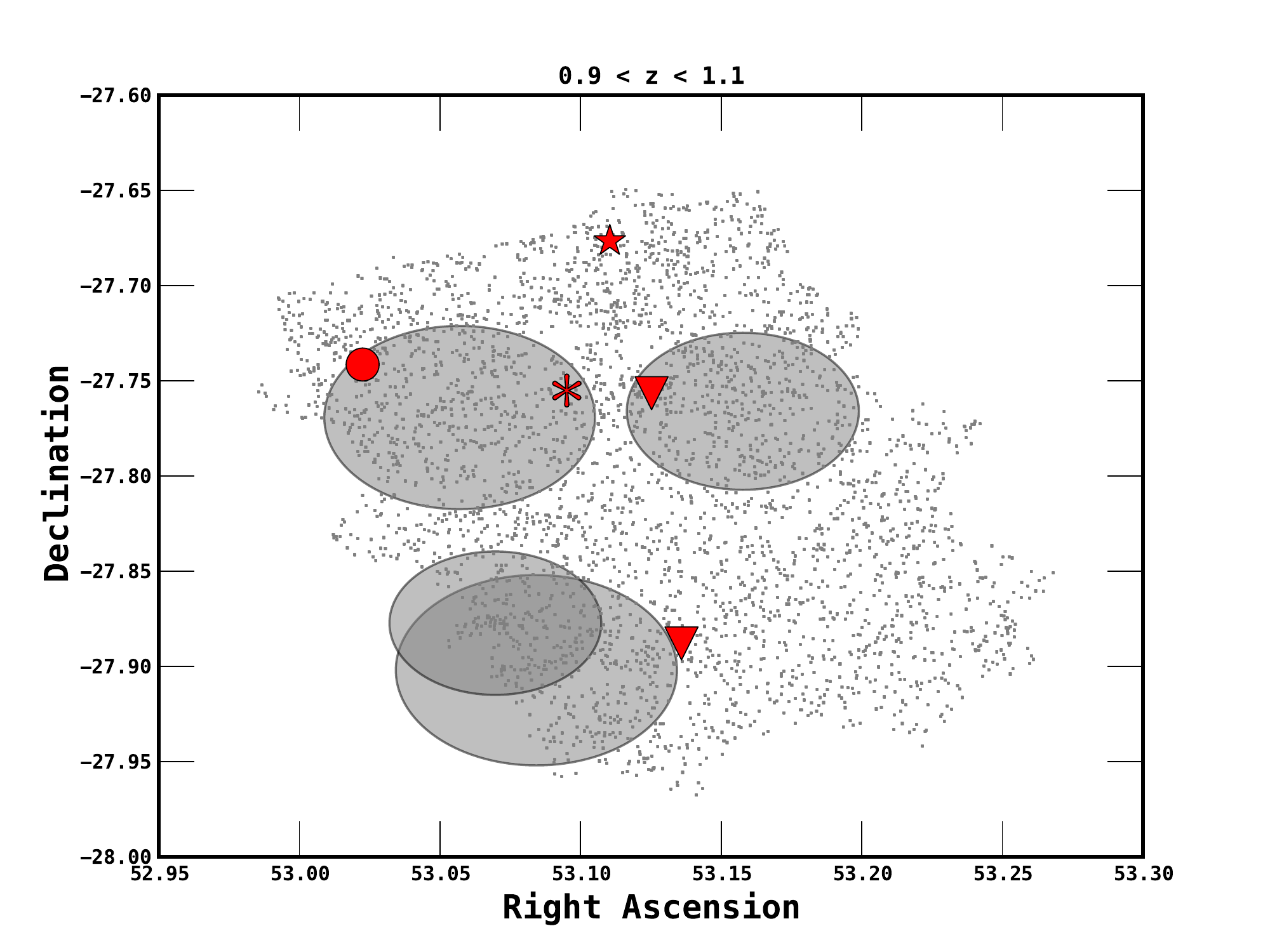}
\includegraphics[width=8cm]{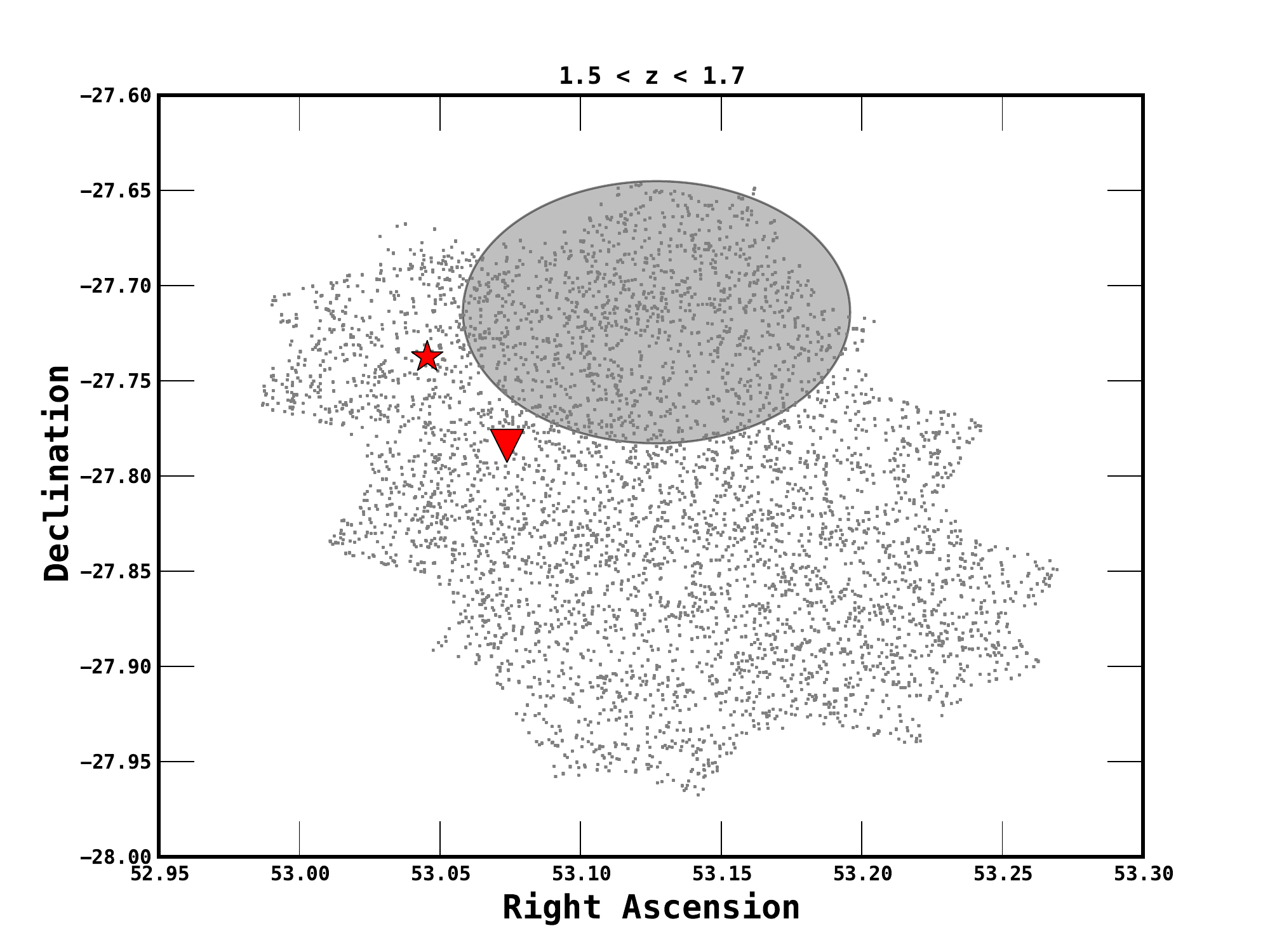}
\includegraphics[width=8cm]{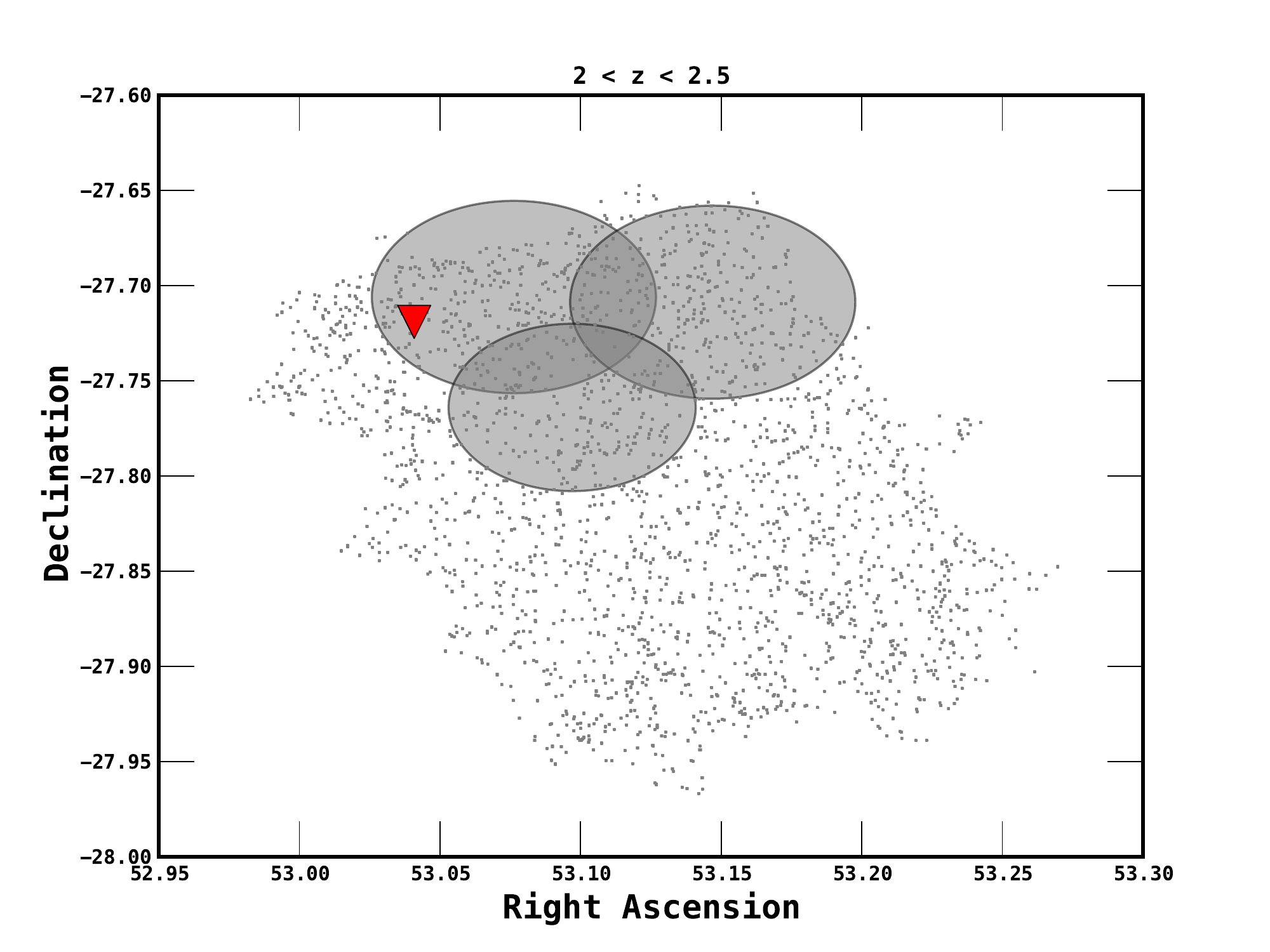}
\caption{The location of variability selected AGN with respect to overdensities listed by Salimbeni et al. (2009), in a number of redshift slices. Only sources with successful fits that have not been identified as false positives are shown.}
\label{F:clusters}
\end{figure*}

To asses the environment of our AGN sample, we use the large-scale structure maps for GOODS-S from \cite{salimbeni_comprehensive_2009}. They used the catalogues of galaxies in GOODS-S with both spectroscopic and photometric redshifts to identify groups and clusters of galaxies. A total of twelve over-densities where found in GOODS-S, with redshifts between $\sim$0.6--2.5. These over-densities overall represent larger groups rather than clusters, with virial masses on the order of $10^{14} M_\odot$ and peak overdensities of 6--10. Fig. \ref{F:clusters} shows the galaxy groups from \cite{salimbeni_comprehensive_2009} as well as all variability selected AGN in the same redshift regime.

It is noticeable that some sources that are located near the centres of over-densities have been classified as late-type or disk-bulge mixtures while interacting systems as well as luminous AGN are either located towards the edges of over-densities or in the field areas. This seems to run counter to previous findings and theoretical expectations \citep[e.g.]{hopkins_characteristic_2009,strand_agn_2008,hickox_host_2009} that high luminosity systems show enhanced clustering on small scales while Seyferts do not. In particular, this is surprising in  the light of the expected connection of quasars with major mergers and Seyferts with secular processes \citet{hopkins_characteristic_2009}. Generally, our sources do not seem to favour dense areas, somewhat consistent with finding that AGN are on average slightly less clustered than normal galaxies \cite[e.g.][]{constantin_clustering_2006,li_clustering_2006}.

\begin{figure}
\includegraphics[width=8cm]{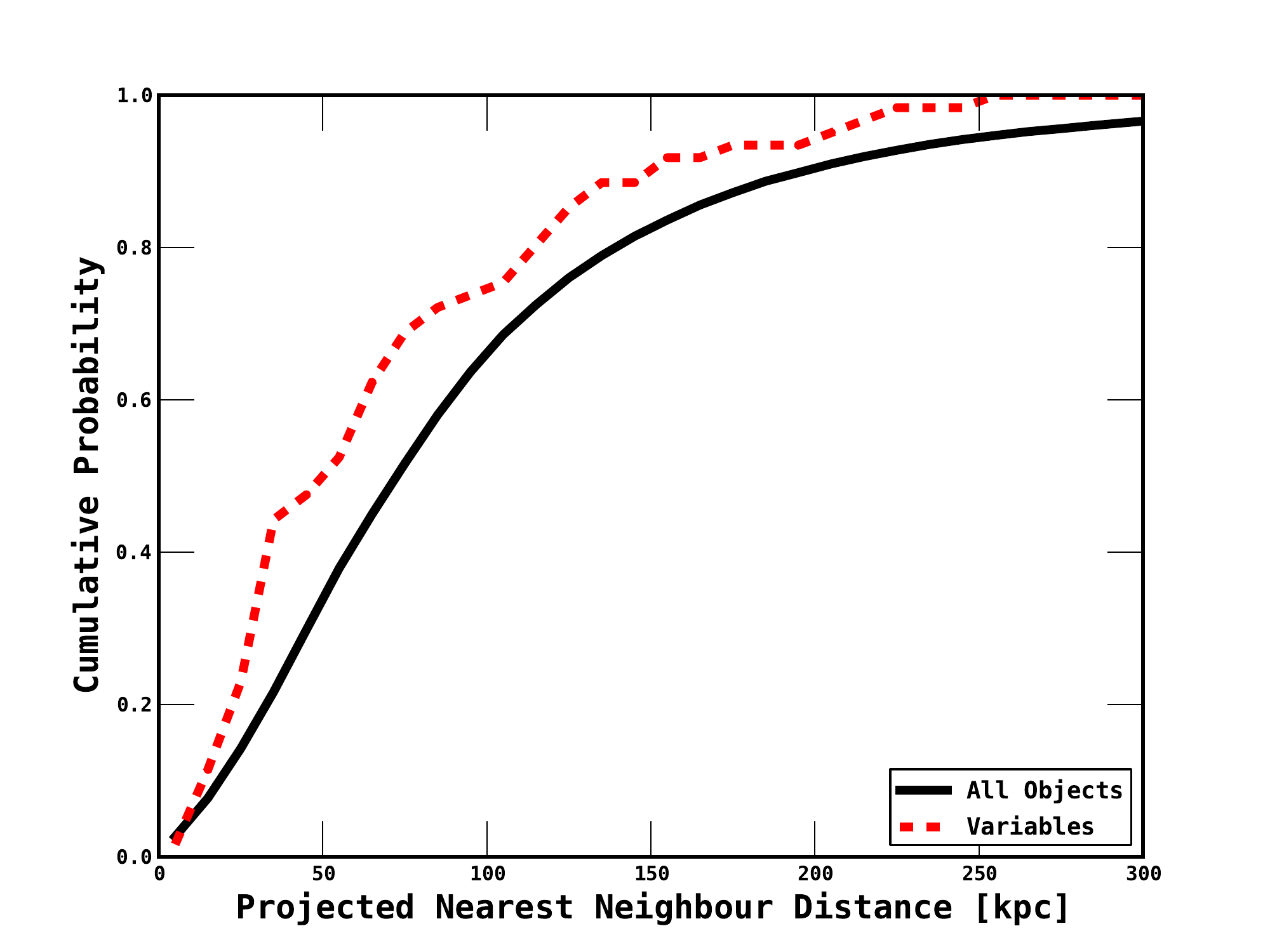}
\caption{Cumulative histograms for nearest projected nearest neighbour distances for variable AGN and control sample respectively.}
\label{F:neighbors}
\end{figure}

Another hint towards triggering mechanisms are nearest neighbour distances. We measure the projected distance to the nearest neighbour for each object and compare the results for the AGN to those of the general population. The distributions differ significantly: variability selected sources show a clear excess of close-by neighbours (Fig. \ref{F:neighbors}). The significance of the nearest neighbour excess depends on the accepted redshift mismatch used for matching a neighbour.  (The values are: p=$4.5\times 10^{-5}$ for $\sigma_{z}=0.1$, p=0.001 for $\sigma_{z}=0.05$, and p=0.08 for $\sigma_{z}=0.01$ using 2 sample KS). The decreasing significance for stricter matching in redshift might indicate a spurious result. However, with a typical scatter in the photometric redshifts of about 0.04, the more strict matching criteria will likely miss actual close neighbours. We find that using the 0.05 results (corresponding to typical scatter in the photometric redshifts) the closest neighbours of AGN hosting galaxies are on average at a projected distance of about 72kpc, while those of the control sample are on average at a distance of about 101kpc (Fig. \ref{F:neighbors}). The results argue for a possible enhancement of AGN activity through tidal interactions. There is no clear trend for galaxies with more close-by neighbours to preferentially have disturbed morphologies or star-burst SEDs. This might indicate that weak tidal interaction may trigger AGN activity without significantly disturbing the host. The implications of these finding will be discussed in more detail in Section \ref{S:discussion}.

\section{Discussion}
\label{S:discussion}

In this paper, we have shown that variability selection identifies low luminosity AGN up to moderately high redshift. We have also demonstrated that commonly used selection methods would have missed a large number of these objects. We have also clearly demonstrated that those sources are indeed AGN using SED fitting. In this section, we aim to discuss the implications of these findings for AGN triggering mechanisms.

Theoretical models propose a number of different possible triggering mechanisms for AGN. Triggering an AGN requires funnelling substantial amounts of cold gas to the vicinity of the black hole while loosing a large fraction of its angular momentum. Several mechanisms could be invoked to achieve this.

The most popular triggering mechanisms currently discussed is major (wet) mergers: during a major merger of two gas rich systems, a large reservoir of gas is available, gravitational torques allow cold gas to settle to the centre. This gas reservoirs can be used to fuel both a central starburst and an AGN. This scenario has been proposed early \citep{sanders_luminous_1996,silk_quasars_1998} and has recently gained popularity due to its ability to alleviate problems in galaxy formation models \citep[e.g.][]{hopkins_cosmological_2008,somerville_semi-analytic_2008}. However, while early observations seemed to support this scenario \citep[e.g.][]{canalizo_quasi-stellar_2001}, recent studies seem to imply that it is not responsible for the triggering of a majority of AGN \citep{cisternas_bulk_2011,cisternas_secular_2011,kocevski_candels:_2011,schawinski_heavily_2012}.

Observable signatures for an AGN triggered in a recent merger are disturbed morphology, a recent star-burst as well as a possible lack of close-by neighbours. While disturbed morphologies are common in the host galaxies of the variability selected AGN, they are seldom accompanied by star-burst SEDs. Also, the fact that the closest neighbour of the AGN in this sample are at a smaller distance than for the general population might speak against such a scenario. Also, the host galaxies of the AGN in this sample do not preferentially populate the green valley, as is expected for many models \cite[e.g.][]{hopkins_cosmological_2008}. However, since we did not attempt to compare the morphologies and star formation rates to a properly matched comparison sample, these results remain tentative. While we cannot exclude major mergers as a trigger for at least some of the AGN in this sample, we do not find evidence for major mergers being the dominant process of AGN triggering. Our findings agree with recent studies finding that while merger triggering is likely in a subset of AGN \citep{ramos_almeida_are_2011}, it is not dominant in the general population \citep{kocevski_candels:_2011,cisternas_bulk_2011,cisternas_secular_2011,schawinski_heavily_2012}. On the other hand, the large incidence of disturbed morphology in our sample are consistent with other findings showing that disturbed morphologies are common in low-luminosity AGN \citep{de_robertis_ccd_1998,bennert_evidence_2008}.

Minor mergers occur much more frequently than major mergers \citep{croton_many_2006,somerville_semi-analytic_2008,lotz_major_2011} but their signatures are more difficult to identify: disturbed morphologies last for shorter time in minor mergers and the morphologies deviate less from those of undisturbed galaxies \citep{lotz_major_2011}. Theoretical models suggest that minor mergers might indeed be important triggering mechanisms for lower luminosity AGN \citep[e.g.][and references therein]{hopkins_characteristic_2009}. Minor mergers can cause disturbed morphologies in the host \citep{lotz_major_2011} but are likely not sufficient to trigger major starbursts \cite{somerville_semi-analytic_2008,hopkins_characteristic_2009}. This scenario agrees well with the findings that low luminosity AGN show disturbed morphology but no signs of starbursts. This also is broadly consistent with the properties of  Seyfert host galaxies \citep{de_robertis_ccd_1998,bennert_evidence_2008,constantin_active_2008}.

Other triggering mechanisms have also been suggested, including cold flows \citep{bournaud_black_2011}, tidal triggering through first passage before a merger \citep{hopkins_cosmological_2008} and bars in spiral galaxies\citep[e.g.][]{shlosman_bars_1989,knapen_subarcsecond_2000,alexander_what_2011}. Given the dominance of $z<1-2$ sources in our sample, cold flow triggering is unlikely for sources in this particular sample since cold flows are more common at higher redshifts \citep{dekel_galaxy_2006}. Also, the process proposed in \cite{bournaud_black_2011} will predominantly produced rather heavily obscured AGN, which are not part of our sample due to the selection technique in this study.  Spiral bars are not addressed in this study, the resolution is not sufficient to clearly examine the presence of bars in this sample, and we will therefore not comment on this possibility. However, we do note that late-type galaxies only represent a small subset of our sample at low redshift.

A mechanisms that seems possible for our sample are triggering during first passage. We find an excess of close neighbours in our sample compared to the general galaxy population, but no clear difference in the general density. This is consistent with first passage triggering. Our findings are also consistent with other studies finding enhanced AGN fractions in close pairs \citep[e.g.][]{ellison_galaxy_2011,farina_study_2011,koss_understanding_2012}. Interestingly, obscured low luminosity AGN are found to have more close neighbours than unobscured sources \citep{dultzin-hacyan_close_1999,koulouridis_local_2006,de_robertis_ccd_1998-1}. This might indicate that first passage triggering is connected with higher levels of obscuration. Given the preference of variability selection to find unobscured sources, the excess in close neighbours we find might therefore be a lower limit to the actual incidence of AGN with close neighbours.

While this seems somewhat cynical, there are several studies suggesting that AGN activity has no preference at all in environment
\citep[e.g.][]{miller_environment_2003,mclure_cluster_2001,wold_radio-quiet_2001} or host galaxy mass as well as stellar populations \citep{aird_primus:_2011}. These finding would imply a 'random' triggering process that has no preference on galaxy properties or environment. We caution therefore that this study is hampered by a lack of comparison sample, other studies have demonstrated that well matched comparison samples are essential since some apparent trends can be solely due to selection effects \citep{kocevski_candels:_2011,cisternas_bulk_2011,cisternas_secular_2011,aird_primus:_2011}.

Our findings are inconsistent with major merger triggering in a majority of sources, especially at the low luminosity end. Our data suggest that minor mergers and triggering through first passage likely play a large role for lower luminosity AGN. A more detailed analysis of triggering mechanisms will require carefully matched control samples, this is however beyond the scope of the current study.

\section{Summary and Conclusions}
\label{S:summary}

In this paper we have presented the SEDs, host galaxy properties and environmental properties of variability selected AGN in GOODS-S. Our results can be summarized as follows:

\begin{itemize}
\item It has been demonstrated that variability selection can complement and expand other AGN selection techniques and expand them to much lower  luminosities. Variability selection can reveal AGN too faint to be detected in even extremely deep X-ray exposures as well as AGN in which the galaxy emission is too dominant to reveal them as AGN in IR colour-colour plots. We have also shown that variability selection indeed identifies low-luminosity AGN and that contamination with normal galaxies is extremely small (6/88), consistent with expected false positive rates. This demonstrates that the method used in \citet{villforth_new_2010} predicts expected false positive rates correctly.
\item Careful SED analysis has revealed that a considerable number of red point-source in the variability selected sample turn out to be stars, in particular of types G-M. In total 24/88 objects from the initial sample turn out to be stars. This reveals the possibility of using variability selection also to find possible rare halo stars as well as the importance of more carefully checking point sources for further variability selection studies.
\item Of the 61 AGN candidates for which SED fits were performed, six are likely false positives, in good agreement with the expected number of false positives in the sample. Of the 55 confirmed AGN, 43 had successful fits, leaving 12 failed fits. Twenty sources have X-ray detection, five have radio detection, three are detected in both X-ray and radio. Of the 43 sources with successful fits, ten are quasar dominated, 19 are mixture objects and 14 are galaxy dominated.
\item SED Fits reveal that variability selection is suitable for selecting AGN that contribute as little as 10\% to the overall emission in a given bands. Variability selection has revealed a considerable number of considerable sub-Seyfert AGN up to redshifts close to 3.
\item Star formation in the hosts of AGN appears most prominent in the highest luminosity objects, while lower luminosity objects appear quiescent or moderately star-forming.
\item The AGN hosts in these study do not have a preferred location with respect to the red sequence and blue cloud, rather the lower redshift low luminosity objects are located in the red sequence and partially the green valley, while higher redshift and higher luminosity objects tend to lie in the blue cloud and partially the green valley. It is not possible to determine for this sample if the redshift or luminosity are the deciding factor for this difference.
\item Disturbed morphologies are common in variability selected AGN and dominate the sample at low AGN luminosities, elliptical galaxies as hosts are rare.
\item We find that the luminosity of AGN and their hosts are generally correlated, but that the AGN luminosity rises faster than that of its host. These finding indicate that either the Eddington ratio or black hole mass to galaxy mass ratio might be higher in more luminous AGN.  There are also tentative findings that this trend is stronger for AGN with ongoing starbursts and that AGN with disturbed morphologies on average have higher Eddington ratios or comparably overmassive black holes given their host galaxy mass.
\item There is no clear evidence for AGN favouring either very high density or very low density environments, however, there is a trend for AGN near the cores of groups to be hosted by late-type galaxies, high luminosity AGN are only found in the field.
\item Nearest neighbours are about 25\% closer in the variability selected AGN sample than they are in the general population, arguing for tidal interactions as possible triggers for weak AGN activity.
\item We find possible indications for different triggering mechanisms at the high and low luminosity end. Lower luminosity AGN in this sample appear to be connected to events which disturb the hosting galaxy but are not capable of triggering extreme star-bursts. We suggest minor mergers or tidal forces during first passage as possible triggering mechanisms. Higher luminosity AGN seem to be connected to events triggering strong starbursts, widely consistent with major mergers. However, since the AGN inhibits a careful analysis of the host galaxy morphology in those systems, it is unclear if they are connected to major mergers or other possible triggers. We would like to caution that we have not properly analysed selection effects. Further careful comparison with a control sample of non-AGN hosting galaxies could show if these findings are due to selection effects. This is however beyond the scope of the current paper.
\end{itemize}

This study therefore demonstrated the strength of variability selection, for the first time determined its relaibility in identifying low luminosity AGN that cannot be found and gave preliminary hints for the possible AGN triggering mechanisms in low luminosity high redshift AGN.

\section*{Acknowledgements}
We would like to thank the anonymous referee for a careful reading of the manuscript and constructive comments. We would like to acknowledge funding through National Science Foundation (NSF) Award AST-1009628 as well as HST grants AR-09935, GO-10134, GO-10530 and GO-11262.

\bibliographystyle{mn2e}
\bibliography{VariableAGN}

\appendix

\section{Morphological Classifications}
\label{A:morphology}

Detailed Morphological Classifications are provided in Table \ref{T:AGN}. All Morphological classifications are performed by eye in the $z$ band, additionally, we also inspected other available HST bands (corresponding to approximately U,V and I bands) to verify our classification. The resolution of the data is 0.1 arcsec, corresponding to a physical scale between ~0.2kpc (z=0.1) and 0.8kpc (z=3). It should be noted that by using the $z$ band as the prime band for inspection, we introduce possible biases since shorter and shorter rest-wavelengths are sampled at higher redshifts. In particular, this might cause us to overestimate the incidence of disturbed morphologies at higher redshifts.

Unresolved sources can either be classified as 'true unresolved' (U), i.e. visual classification does not show signs of host galaxy contribution and 'weak host' (UH) in which visual classification shows clear signs of host galaxy contribution. For those sources, additional flags can be set (see below).

For the extended sources, we have five main morphology classes as well as additional flags (see below). The five main are: late-type galaxies (D, clear disk, no strong bulge), early-type galaxies (E), 'spheroidal': a clear disk-bulge component (S), interacting/disturbed (I), all other sources that are too faint for a clear classification are classified as unclear (X). The following flags are available additionally for all extended as well as unresolved with clear host sources: clear dominant core (C), nearby neighbour (N), edge-on late-type (E), face-on late-type (F), clear merger (M), weak tidal tails (T), multiple cores/clumpy/trainwreck (X).

\begin{table*}
\centering
\begin{minipage}{180mm}
\caption{Variability Selected AGN. Name: GOODS ID; RA: Right Ascension; Dec: Declination; $mag_{z}$: $z$ band apparent magnitudes; z: redshift, superscript indicates if the redshift is spectroscopic (s) or photometric (p, from \protect \cite{dahlen_detailed_2010}); morphology: morphological classification, see Section \ref{A:morphology} for details; type: spectral type, Quasar Dominated (Q), Mixture (M), Galaxy Dominated (G), Failed Fit (X), False Positive (FP); AGN: Best Fit AGN SED; Galaxy: Best Fit Galaxy SED; ratio: AGN contribution; Xray?: x-ray detected, yes/no; Radio?: radio detected, yes/no.}
\begin{tabular}{lcccccccccccc}
\hline
Name & RA (J2000) & Dec (J2000) & $mag_{z}$ & z & Morphology & Type & AGN & Galaxy & Ratio & Xray? & Radio?\\
\hline
J033203.00-274213.6 & 53.01248 & -27.7037872 & 25.16 & 2.66$^{p}$  & EX & Q & BQSO1 & Spi4 & 0.9 & No & No \\
J033203.01-274544.7 & 53.0125331 & -27.7624232 & 25.12 & 0.54$^{p}$  & ED & G & BQSO1 & Spi4 & 0.1 & No & No \\
J033203.26-274530.3 & 53.013574 & -27.7584257 & 22.72 & 0.04$^{p}$  & ED & G & BQSO1 & Spi4 & 0.1 & No & No \\
J033205.40-274429.2 & 53.0224983 & -27.7414518 & 22.99 & 1.039$^{s}$  & EDF & G & BQSO1 & Spi4 & 0.05 & No & No \\
J033208.68-274508.0 & 53.036184 & -27.7522317 & 22.71 & 1.296$^{s}$  & EDC & X & - & - & - & No & No \\
J033209.57-274634.9 & 53.0398585 & -27.7763722 & 22.97 & 0.5375$^{s}$  & EEC & FP & - & - & - & No & No \\
J033209.58-274241.8 & 53.0399142 & -27.7116218 & 23.44 & 0.681$^{s}$  & EIC & G & BQSO1 & Spi4 & 0.1 & No & No \\
J033209.80-274308.6 & 53.0408277 & -27.7190613 & 23.15 & 2.48$^{p}$  & EIM & Q & BQSO1 & Ell2 & 0.95 & No & No \\
J033210.52-274628.9 & 53.0438334 & -27.7747007 & 23.88 & 1.615$^{s}$  & EXN & FP & - & - & - & No & No \\
J033210.91-274414.9 & 53.0454704 & -27.7374846 & 22.34 & 1.613$^{s}$  & UQXRH & Q & Mrk231 & - & 1.0 & Yes & Yes \\
J033211.02-274919.8 & 53.045918 & -27.8221721 & 23.45 & 1.9431$^{s}$  & EIMX & M & BQSO1 & Spi4 & 0.3 & No & No  \\
J033212.16-274408.8 & 53.0506774 & -27.7357915 & 24.81 & 0.74$^{p}$  & EX & M & BQSO1 & Ell2 & 0.3 & No & No \\
J033213.21-274715.7 & 53.0550544 & -27.7876915 & 24.19 & 0.523$^{s}$  & EDEC & G & BQSO1 & Ell2 & 0.1 & No & No \\
J033217.06-274921.9 & 53.0710682 & -27.8227401 & 19.19 & 0.34$^{s}$  & EDT & G & TQSO & S0 & 0.1 & Yes & No \\
J033217.14-274303.3 & 53.0714326 & -27.7175864 & 20.58 & 0.566$^{s}$  & EE & M & TQSO & S0 & 0.2 & Yes & Yes\\
J033217.72-274703.0 & 53.0738469 & -27.7841607 & 23.63 & 1.607$^{s}$  & EI & Q & BQSO1 & Sdm & 0.9 & No & No \\
J033218.24-275241.4 & 53.0760024 & -27.8781606 & 24.26 & 2.801$^{s}$  & UQX & Q & BQSO1 & Sey18 & 0.9 & Yes & No \\
J033218.70-275149.3 & 53.0778965 & -27.8637054 & 21.76 & 0.458$^{s}$  & ESE & M & QSO1 & Ell5 & 0.15 & No & No \\
J033218.81-274908.5 & 53.0783542 & -27.8190409 & 23.88 & 1.128$^{s}$  & EIX & X & - & - & - & No & No \\
J033218.84-274529.2 & 53.078519 & -27.7581146 & 18.79 & 0.296$^{s}$  & EE & G & TQSO & Ell13 & 0.05 & Yes & No \\
J033219.81-275300.9 & 53.0825539 & -27.8835727 & 24.50 & 3.706$^{s}$  & EI & X & - & - & - & No & No \\
J033219.86-274110.0 & 53.0827492 & -27.686119 & 23.40 & 1.542$^{s}$  & EIM & FP & - & - & - & No & No \\
J033222.82-274518.4 & 53.0950956 & -27.7550986 & 23.24 & 1.087$^{s}$  & EXC & M & BQSO1 & Ell2 & 0.3 & No & No \\
J033223.53-274707.5 & 53.0980434 & -27.785425 & 23.26 & 0.69$^{p}$  & EDCF & M & BQSO1 & Ell2 & 0.2 & No & No \\
J033224.23-274129.5 & 53.1009433 & -27.691518 & 24.89 & 4.34$^{p}$  & EI & Q & BQSO1 & Ell2 & 1.0 & No & No \\
J033224.54-274010.4 & 53.1022685 & -27.6695645 & 21.91 & 0.96$^{p}$  & EI & X & - & - & - & Yes & No \\
J033224.80-274617.9 & 53.1033449 & -27.7716431 & 23.09 & 1.306$^{s}$  & EX & X & - & - & - & No & No \\
J033225.10-274403.2 & 53.1045636 & -27.7342138 & 17.76 & 0.076$^{s}$  & ES & FP & QSO1 & S0 & 0.0 & No & No \\
J033225.99-274142.9 & 53.1082952 & -27.6952603 & 24.56 & 0.0459$^{s}$  & EICN & Q & BQSO1 & Spi4 & 0.95 & No & No \\
J033226.40-275532.4 & 53.109991 & -27.9256529 & 24.44 & 1.77$^{p}$  & EI & X & - & - & - & No & No \\
J033226.49-274035.5 & 53.1103938 & -27.6765399 & 19.62 & 1.031$^{s}$  & UQXHDCT & Q & QSO1 & I19524 & 0.9 & Yes & No \\
J033227.01-274105.0 & 53.1125287 & -27.6847238 & 19.03 & 0.734$^{s}$  & UQ & M & QSO1 & S0 & 0.75 & Yes & No \\
J033227.18-274416.5 & 53.113269 & -27.73791 & 19.38 & 0.418$^{s}$  & ES & G & TQSO & S0 & 0.05 & Yes & No \\
J033227.51-275612.4 & 53.114644 & -27.9367764 & 24.04 & 0.663$^{s}$  & EXC & X & - & - & - & No & No \\
J033228.30-274403.6 & 53.1179209 & -27.7343234 & 23.69 & 3.25$^{p}$  & EI & X & - & - & - & Yes & No \\
J033229.88-274424.4 & 53.1244949 & -27.7401248 & 16.46 & 0.076$^{s}$  & EDCF & G & N6240 & - & - & No & Yes \\
J033229.98-274529.9 & 53.1249148 & -27.7583013 & 21.11 & 1.215$^{s}$  & UQ & M & TQSO & I22491 & 0.5 & Yes & No \\
J033229.99-274404.8 & 53.1249588 & -27.7346753 & 16.85 & 0.076$^{s}$  & EDCF & M & Sey2 & Spi4 & 0.3 & Yes & Yes \\
J033230.06-274523.5 & 53.1252547 & -27.756535 & 21.82 & 0.955$^{s}$  & EIX & M & TQSO & S0 & 0.3 & Yes & No \\
J033230.22-274504.6 & 53.1258995 & -27.7512749 & 21.51 & 0.738$^{s}$  & UQ & M & BQSO1 & Sc & 0.5 & Yes & No \\
J033230.36-275133.2 & 53.1264805 & -27.8592312 & 25.02 & 0.59$^{p}$  & EX & X & - & - & - & No & No \\
J033232.04-274523.9 & 53.1334996 & -27.7566329 & 23.44 & 0.43$^{p}$  & EIX & M & BQSO1 & Ell2 & 0.15 & No & No \\
J033232.49-275044.0 & 53.1353687 & -27.8455431 & 23.44 & 0.41$^{p}$  & EDC & G & QSO1 & Ell2 & 0.1 & No & No \\
J033232.61-275316.7 & 53.1358826 & -27.8879636 & 22.96 & 0.988$^{s}$  & EIMC & G & BQSO1 & Spi4 & 0.1 & No & No \\
J033232.67-274944.6 & 53.1361117 & -27.829048 & 18.29 & 0.14$^{p}$  & ES & G & BQSO1 & S0 & 0.05 & Yes & No \\
J033233.68-274035.6 & 53.1403383 & -27.6765489 & 24.97 & 0.81$^{p}$  & EX & X & - & - & - & No & No \\
J033235.38-274704.3 & 53.1474321 & -27.7845155 & 25.16 & 0.91$^{p}$  & EX & X & - & - & - & No & No \\
J033236.92-275308.4 & 53.1538193 & -27.885679 & 24.42 & 0.42$^{p}$  & EXT & M & BQSO1 & Ell2 & 0.3 & No & No \\
J033238.12-273944.8 & 53.1588307 & -27.6624444 & 20.45 & 0.837$^{s}$  & UQ & Q & BQSO1 & I19524 & 0.95 & Yes & No \\
J033238.89-275406.7 & 53.1620545 & -27.9018501 & 24.62 & 1.467$^{s}$  & EXN & M & BQSO1 & Ell2 & 0.8 & No & No \\
J033239.09-274601.8 & 53.1628593 & -27.7671602 & 21.02 & 1.216$^{s}$  & UQX & M & BQSO1 & I19524 & 0.5 & Yes & No \\
J033239.47-275300.5 & 53.1644567 & -27.8834689 & 20.55 & 0.64$^{p}$  & EENC & G & - & S0 & 0.0 & No & Yes \\
J033240.89-275449.2 & 53.1703678 & -27.9136643 & 25.19 & 2.53$^{p}$  & EX & Q & BQSO1 & Sdm & 0.9 & No & No \\
J033241.87-274651.1 & 53.1744478 & -27.7808655 & 23.35 & 0.7$^{p}$  & EI & FP & - & Spi4 & 0.0 & No & No \\
J033243.24-274914.2 & 53.1801493 & -27.8206046 & 22.70 & 1.906$^{s}$  & UQX & M & TQSO & I22491 & 0.5 & Yes & No \\
J033243.93-274351.1 & 53.1830401 & -27.7308524 & 24.28 & 0.63$^{p}$  & UQ & M & BQSO1 & Ell2 & 0.5 & No & No \\
J033245.02-275207.7 & 53.1875887 & -27.8688155 & 22.52 & 0.721$^{s}$  & EDCT & G & Mrk231 & S0 & 0.05 & No & No \\
J033246.37-274912.8 & 53.1932061 & -27.8202112 & 21.95 & 0.683$^{s}$  & ES & FP & TQSO & S0 & 0.05 & No & No \\
J033247.98-274855.7 & 53.1999289 & -27.8154702 & 20.57 & 0.233$^{s}$  & EE & X & - & - & - & Yes & No \\
J033252.88-275119.8 & 53.2203537 & -27.8555099 & 21.87 & 1.22$^{s}$  & EDC & M & TQSO & N6090 & 0.3 & Yes & No \\
J033253.44-275001.4 & 53.2226606 & -27.8337103 & 24.35 & 1.245$^{s}$  & EIX & M & BQSO1 & Spi4 & 0.8 & No & No \\
\hline
\end{tabular}
\label{T:AGN}
\end{minipage}
\end{table*}

\begin{table*}
\centering
\begin{minipage}{140mm}
\caption{Stars in Variability Selected Sample. Spectral Types Notes: a: This work; b: Proper Motion Detected; c: \protect \cite{popesso_great_2009}; d: \protect \cite{wolf_calibration_2008}; e: \protect \cite{groenewegen_eso_2002}; f: \protect \cite{santini_star_2009}; h: \protect \cite{taylor_public_2009}; i: star has close neighbour, spectral type not secure. }
\begin{tabular}{lcccc}
\hline
Name & Right Ascension (J2000) & Declination (J2000) & Spectral Type & $mag_{z}$ [AB] \\
\hline
J033204.41-274635.5 &   53.018367 & -27.7765167 & M $^{a}$     &  21.14 \\
J033205.11-274317.5 &   53.021302 & -27.7215411 & M  $^{a}$   &  21.91 \\
J033213.34-274210.5 &  53.0555659 & -27.7029202 & M $^{a}$    &  23.84 \\
J033215.16-274754.6 &  53.0631513 & -27.7985117 & K  $^{a}$   &  22.98 \\
J033215.93-275329.3 &  53.0663591 & -27.8914797 & M2Iab $^{c}$&  19.86 \\
J033216.34-274851.7 &  53.0681031 & -27.8143745 & M $^{a}$    &  21.98 \\
J033216.87-274916.7 &  53.0702817 & -27.8212937 & K  $^{a}$   &  23.40 \\
J033220.80-275144.5 &  53.0866794 & -27.8623513 & M  $^{a}$   &  21.22 \\
J033221.52-274358.7 &  53.0896509 &  -27.732984 & M2Iab $^{d}$ &  21.00\\
J033225.15-274933.3 &  53.1048118 & -27.8259053 & G0Iab $^{c}$ &  21.61\\
J033227.86-275335.6 &  53.1160832 & -27.8932186 & M3III $^{e,i}$ &  21.84\\
J033227.87-275335.9 &  53.1161233 & -27.8933035 & M3III $^{e,i}$ &  21.63\\
J033228.45-274203.8 &   53.118527 & -27.7010451 & K $^{a}$    &  24.23\\
J033231.80-275110.4 &  53.1324925 & -27.8528853 & M3III $^{d,i}$ &  21.03\\
J033231.82-275110.6 &  53.1325905 & -27.8529435 & M3III $^{d,i}$ &  21.06\\
J033231.94-274531.3 &  53.1330882 & -27.7587052 & G5V $^{f}$   &  21.11\\
J033232.12-275636.8 &  53.1338346 & -27.9435593 & WD $^{a,b}$   &  21.04\\
J033232.32-274316.4 &  53.1346778 & -27.7212189 & M $^{a}$    &   22.11\\
J033237.93-274609.1 &  53.1580245 &  -27.769192 & K  $^{a,b}$   &  19.96\\
J033241.05-275234.1 &  53.1710547 & -27.8761468 & G0V  $^{c}$ &  20.72\\
J033242.61-275453.8 &  53.1775348 & -27.9149453 & G8Iab $^{c}$ &   20.82\\
J033244.10-275212.9 &  53.1837535 & -27.8702564 & K3III $^{d}$ &  20.88\\
J033246.39-274820.1 &  53.1932909 & -27.8055737 & M2Iab $^{h}$ &  21.16\\
J033247.53-275159.9 &  53.1980298 & -27.8666421 & G5V $^{a}$  &  21.01\\
\hline
\end{tabular}
\label{T:Stars}
\end{minipage}
\end{table*}

\label{lastpage}

\end{document}